\documentclass{article}

\scrollmode
\usepackage{amsmath}
\usepackage{amsfonts}
\usepackage{amssymb}
\usepackage{latexsym}
\usepackage{stmaryrd}
\usepackage{array}
\usepackage{exscale}
\newcommand{\nc}{\newcommand}
\newcommand{\ol}{\overline}
\newcommand{\ul}{\underline}
\newcommand{\es}{\emptyset}
\newcommand{\sm}{\setminus}
\newcommand{\ve}{\varepsilon}
\newcommand{\vp}{\varphi}

\newcommand{\bc}{\bigcup}

\newcommand{\Lra}{\Leftrightarrow}

\newcommand{\Ra}{\Rightarrow}

\newcommand{\ra}{\rightarrow}

\newcommand{\sse}{\subseteq}

\newcommand{\spe}{\supseteq}
\newcommand{\fa}{\forall}
\newcommand{\ex}{\exists}
\newcommand{\mr}{\mathrm}
\newcommand{\mc}{\mathcal}
\newcommand{\mf}{\mathfrak}

\newcommand{\DMO}{\DeclareMathOperator}
\newcommand{\DST}{\displaystyle}

\newcommand{\NN}{\mathbb{N}}
\newcommand{\NNZ}{\NN_0}

\newcommand{\RR}{\mathbb{R}}

\newcommand{\PP}{\mathbb{P}}

\newcommand{\TT}{\mathbb{T}}
\usepackage{listings}
\newcommand{\inl}[1]{\lstinline$#1$}
\newcommand{\und}{{\:\wedge\:}} 
\newcommand{\mb}{{\:|\:}} 
\newcommand{\set}[1]{\{ #1 \}}
\newcommand{\setb}[1]{\big \{ \, #1 \, \big \}}
\nc{\simlvi}[1]{\!\sim_{#1}}
\DeclareMathOperator{\rstr}{|} 
\nc{\apprel}[3]{{#1}(#2)_{(#3)}} 
\nc{\cmpli}[1]{\complement^1_{#1}} 
\nc{\cmplzi}[1]{\complement^0_{#1}} 
\nc{\cmplzoi}[1]{\complement^*_{#1}} 

\nc{\zf}{\mr{ZF}}
\nc{\zfmf}{\zf^0} 
\nc{\zfc}{\mr{ZFC}}
\nc{\zfcmf}{\zfc^0} 
\nc{\bst}{\mr{BST}} 
\newcommand{\tb}[2]{\set{#1, \dots, #2}} 
\providecommand{\abs}[1]{\lvert #1 \rvert} 
\providecommand{\norm}[1]{\lVert #1 \rVert} 
\makeatletter
\DeclareRobustCommand{\genericinterval}[2]{%
  \@ifstar{\genericinterval@star{#1}{#2}}{\genericinterval@nostar{#1}{#2}}}
\newcommand{\genericinterval@star}[4]{\mathopen{}\mathclose{\left#1#3,#4\right#2}}
\newcommand{\genericinterval@nostar}[4]{\mathopen{#1}#3,#4\mathclose{#2}}

\makeatother
\nc{\untit}[2]{{#1}^{#2 \downarrow}} 
\nc{\obit}[2]{{#1}^{#2 \uparrow}} 
\nc{\inzEKi}[1]{\mc{I}^{\mr{V}}_{#1}}

\nc{\inzKEi}[1]{\mc{I}^{\mr{E}}_{#1}}

\nc{\adjEi}[1]{\mc{A}^{\mr{V}}_{#1}}

\nc{\BD}[1]{{#1}\text{-}\mr{BD}}

\DeclareMathOperator{\pot}{\PP} 
\nc{\konv}[2]{{#1}[{#2}]} 
\nc{\actpres}[1]{\Phi_{#1}} 
\newcommand{\floor}[1]{\lfloor #1 \rfloor}

\nc{\Prim}{\mc{PR}} 

\nc{\sselr}{\sse^{\mapsto}}
\nc{\sserl}{\sse^{\mapsfrom}}
\nc{\spelr}{\spe^{\mapsto}}
\nc{\sperl}{\spe^{\mapsfrom}}
\nc{\ball}[1]{\mr{B}^{#1}} 
\nc{\oball}[1]{\breve{\mr{B}}^{#1}} 
\nc{\pball}[1]{\dot{\mr{B}}^{#1}} 
\nc{\prr}[1]{\dot{\RR}^{#1}} 
\nc{\sph}[1]{\mr{S}^{#1}} 
\nc{\ssim}[1]{s\sigma_{#1}} 
\nc{\koerper}[1]{\norm{#1}}
\nc{\Ccovdim}{\mc{CD}}
\nc{\Cinddim}{\mc{SID}}

\nc{\CInddim}{\mc{LID}}

\DeclareMathOperator{\diffop}{D} 
\DeclareMathOperator*{\diffoplimit}{D} 
\nc{\diffopc}[1]{\sideset{_{#1}}{}\diffoplimit} 
\nc{\diffopp}[1]{\diffop_{#1}} 
\nc{\diffopcp}[2]{\sideset{_{#2}}{_{#1}}\diffoplimit} 
\nc{\meanH}[2]{\mf{M}_{#1,#2}} 
\nc{\emean}[2]{\mf{M}_{\exp_{#1},#2}} 
\DeclareMathOperator{\mor}{Mor}
\DeclareMathOperator{\Hom}{Hom} 
\nc{\autoerw}[1]{\mr{Aut}^{#1}} 
\nc{\komma}[2]{(#1 \downarrow #2)} 
\nc{\Kmat}{\mf{MAT}} 
\nc{\Khmat}{\mf{HMAT}} 
\nc{\homfun}[1]{\mor_{#1}(-_1,-_2)} 
\nc{\homfunae}[1]{\mor_{#1}(-_1)} 
\nc{\homfunbe}[1]{\mor_{#1}(-_2)} 
\nc{\homfunxy}[3]{\mor_{#1}(#2(-_1), #3(-_2))}
\nc{\homfunx}[2]{\mor_{#1}(#2(-_1), -_2)}
\nc{\homfuny}[2]{\mor_{#1}(-_1, #2(-_2))}
\nc{\homfuna}[2]{\mor_{#1}(#2, -)} 
\nc{\homfunb}[2]{\mor_{#1}(-, #2)} 
\nc{\hhomfuna}[2]{\Hom_{#1}(#2, -)} 
\nc{\hhomfunb}[2]{\Hom_{#1}(-, #2)} 
\newcommand{\Va}{\mc{V\hspace{-0.1em}A}}

\newcommand{\Lit}{\mc{LIT}}
\newcommand{\Cl}{\mc{CL}}
\newcommand{\Cls}{\mc{CLS}}

\newcommand{\Pcls}[1]{#1\mbox{--}\Cls}

\newcommand{\Pass}{\mc{P\hspace{-0.32em}ASS}}
\newcommand{\Tass}{\mc{T\hspace{-0.35em}ASS}}
\newcommand{\Sat}{\mc{SAT}}

\newcommand{\Usat}{\mc{USAT}}

\newcommand{\Musat}{\mc{M\hspace{0.8pt}U}} 
\newcommand{\Musati}[1]{\Musat_{\!#1}} 
\newcommand{\Smusat}{\mc{S}\Musat} 
\newcommand{\Smusati}[1]{\Smusat_{\!#1}}

\nc{\Clsoo}{\Cls^{1,1}} 
\DeclareMathOperator{\lit}{lit}
\DeclareMathOperator{\var}{var}

\newcommand{\Clash}{\mc{HIT}} 

\newcommand{\Uclash}{\mc{U}\Clash} 
\newcommand{\Uclashi}[1]{\Uclash_{\!\!#1}}

\newcommand{\Ho}{\mc{HO}} 
\newcommand{\Rho}{\mc{R}\Ho} 

\DeclareMathOperator{\res}{\diamond} 

\DeclareMathOperator{\comp}{Comp} 
\DeclareMathOperator{\compr}{\comp_R} 
\DMO{\premr}{F} 
\DMO{\concr}{C} 
\DMO{\allcr}{\widehat{F}} 
\DMO{\semspace}{ss} 
\DMO{\treespace}{ts} 

\DeclareMathOperator{\hardness}{hd}
\DMO{\phardness}{phd} 
\DMO{\whardness}{whd} 
\DMO{\hts}{hs} 
\newcommand{\php}{\mathrm{PHP}}
\DeclareMathOperator{\nds}{nds} 
\DeclareMathOperator{\lvs}{lvs} 
\DeclareMathOperator{\nlvs}{\#lvs} 
\DeclareMathOperator{\nnds}{\#nds} 
\DeclareMathOperator{\height}{ht}
\DeclareMathOperator{\depth}{d}

\newcommand{\pab}[1]{\langle #1 \rangle}
\newcommand{\pao}[2]{\langle #1 \ra #2 \rangle}

\nc{\bth}[1]{\langle{#1}\rangle} 
\DMO{\rsub}{r_S} 
\DMO{\rk}{r} 
\DMO{\rki}{r_{\infty}} 
\nc{\rslur}{\xrightarrow{\text{SLUR}}} 
\nc{\rslurs}{\rslur_{\!*}} 
\DMO{\slur}{slur} 
\nc{\Slur}{\mc{SLUR}} 
\nc{\rkslur}[1]{\xrightarrow{\text{SLUR}_{#1}}} 
\nc{\rkslurs}[1]{\rkslur{#1}_{\!*}} 
\nc{\Altsluri}[1]{\Slur(#1)}
\nc{\Altslurstari}[1]{\Slur\text{\textasteriskcentered}(#1)}
\nc{\Canoni}[1]{\mr{CANON}(#1)}
\nc{\rkslurstar}[1]{\xrightarrow{\text{SLUR\textasteriskcentered}#1}} 
\nc{\rkslursstar}[1]{\rkslurstar{#1}_{\!*}} 
\DMO{\slurstar}{\slur\!\text{\textasteriskcentered}}
\nc{\Urefc}{\mc{UC}}
\nc{\Propc}{\mc{PC}}
\nc{\Wrefc}{\mc{WC}} 
\DeclareMathOperator{\wid}{wid} 

\DeclareMathOperator{\vdeg}{vd} 
\DeclareMathOperator{\minvdeg}{\mu\!\vdeg} 
\DMO{\varmvd}{\var_{\minvdeg}} 
\DMO{\nfc}{fc} 
\DMO{\maxnfc}{\nu\!\nfc} 
\newcommand{\OKsolver}{\texttt{OKsolver}}

\nc{\svbf}{\mc{VB}} 
\nc{\svbfs}{\mc{VB}^*} 
\DMO{\potp}{pp} 
\DMO{\potprec}{NM} 
\DMO{\minnonmer}{\mu{}nM} 
\DMO{\varsing}{\var_s} 
\DMO{\varosing}{\var_{1s}} 
\DMO{\varnosing}{\var_{\neg1s}} 
\nc{\Musatns}{\Musat'} 
\nc{\Musatnsi}[1]{\Musati{#1}'}
\nc{\Smusatns}{\Smusat'} 
\nc{\Smusatnsi}[1]{\Smusati{#1}'}
\nc{\Uclashns}{\Uclash'} 
\nc{\Uclashnsi}[1]{\Uclashi{#1}'}
\nc{\tsdp}{\xrightarrow{\text{sDP}}}
\nc{\tsdps}{\tsdp_{\!*}}
\nc{\tosdp}{\xrightarrow{\text{1sDP}}}
\nc{\tosdps}{\tosdp_{\!*}}
\DMO{\sdp}{sDP} 
\DMO{\osdp}{sDP_1} 
\nc{\cflmusat}{\mc{CF}\Musat} 
\nc{\cflmusati}[1]{\mc{CF}\Musati{#1}}
\nc{\cflimusat}{\mc{CFI}\Musat} 
\DMO{\sNF}{sNF} 
\DMO{\eqp}{eqp} 
\DMO{\sgp}{sp} 
\DMO{\singind}{si} 
\DMO{\osingind}{si_1} 
\DMO{\shyp}{svh} 
\DMO{\sdph}{ssh} 
\DMO{\msdph}{mss} 
\DMO{\osdph}{ssh_1} 
\DMO{\mosdph}{mss_1} 
\newcommand{\Mps}{\mathcal{MPS}} 
\DMO{\mps}{mps} 
\DMO{\purec}{puc} 
\DMO{\doping}{D}
\DeclareMathOperator{\primec}{prc} 
\nc{\glue}[4]{\mr{glue}((#1,#2), (#3,#4))} 
\DMO{\fvdglue}{\boxplus} 
\nc{\gluea}[3]{#1 \boxplus_{#3} #2} 
\DMO{\frl}{fl} 
\nc{\Con}{\mr{Con}}
\nc{\Log}{\mr{Log}}
\nc{\Lin}{\mr{Lin}}
\nc{\Pol}{\mr{Pol}}
\nc{\ExL}{\mr{ExL}}
\nc{\ExP}{\mr{ExP}}
\nc{\CTime}{\mr{CTime}}
\nc{\CSpace}{\mr{CSpace}}
\nc{\LTime}{\mr{LTime}}
\nc{\LSpace}{\mr{L}}
\nc{\NLSpace}{\mr{NL}}
\nc{\LinTime}{\mr{LinTime}}
\nc{\LinSpace}{\mr{LinSpace}}
\nc{\PTime}{\mr{P}}
\nc{\PSpace}{\mr{PSpace}}
\nc{\Np}{\mr{NP}}
\nc{\Conp}{\text{coNP}}
\nc{\NPSpace}{\mr{NPSpace}}
\nc{\CoNPSpace}{\mr{coNPSpace}}
\nc{\ELTime}{\mr{ELTime}}
\nc{\ELSpace}{\mr{ELSpace}}
\nc{\EPTime}{\mr{EPTime}}
\nc{\EPSpace}{\mr{EPSpace}}
\nc{\NEPTime}{\mr{NEPTime}}
\nc{\polydelta}[1]{\Delta_{#1}^{\mr P}}
\nc{\polypi}[1]{\Pi_{#1}^{\mr P}}
\nc{\polysigma}[1]{\Sigma_{#1}^{\mr P}}
\nc{\Ph}{\mr{PH}}

\nc{\Dp}{D^P}
\nc{\PllC}[2]{{\text{$\mr{PT}$/$\mr{WK}$}(#1, #2)}} 
\nc{\Nc}{\mr{NC}}
\nc{\Nci}[1]{\Nc^{#1}}
\nc{\Ac}{\mr{AC}}
\nc{\Aci}[1]{\Ac^{#1}}
\nc{\pmodpoly}{P / \mathrm{poly}}
\nc{\Wh}[1]{\mr{W}[#1]} 
\nc{\Rl}{\mr{RL}}
\nc{\coRl}{\mr{coRL}}
\nc{\Rp}{\mr{RP}}
\nc{\coRp}{\mr{coRP}}
\nc{\Zpp}{\mr{ZPP}}
\nc{\Bpp}{\mr{BPP}}
\nc{\Pp}{\mr{PP}}
\nc{\Reach}{\mr{STCON}} 
\nc{\Undreach}{\mr{USTCON}} 
\nc{\Pcol}[2]{\mr{COL}(#1,#2)} 
\nc{\Pscol}[2]{\mr{SCOL}(#1,#2)} 
\nc{\Psorcol}[2]{\mr{SORCOL}(#1,#2)} 
\nc{\Mss}{\mr{MSS}}
\nc{\Key}{\mr{KEY}}
\nc{\Keyi}[1]{\Key_{\!#1}}
\nc{\Nbmss}{N_{\mr{bm}}} 
\nc{\Nbkey}{N_{\mr{bk}}} 
\nc{\Rnb}{N_{\mr{b}}}
\nc{\Rnk}{N_{\mr{k}}}
\nc{\Rnr}{N_{\mr{r}}}

\nc{\Byte}{\mr{B}[8]}
\nc{\QByte}{\mr{B}[4,8]}
\nc{\KByte}{\mc{B}} 
\nc{\RQByte}{\mc{QB}} 

\nc{\ramz}[3]{\mr{ram}_{#1}^{#2}(#3)} 
\nc{\waez}[2]{\mr{vdw}_{#1}(#2)} 
\nc{\gtz}[2]{\mr{grt}_{#1}(#2)} 
\nc{\pdwaez}[2]{\mr{vdw}_{#1}^{\mr{pd}}(#2)} 
\nc{\absfeh}[1]{\delta_{#1}} 
\nc{\relfeh}[1]{\ve_{#1}} 
\usepackage{theorem} 
\usepackage[driverfallback=hypertex]{hyperref}
\newtheorem{defi}{Definition}[section]
\newtheorem{lem}[defi]{Lemma}
\newtheorem{thm}[defi]{Theorem}
\newtheorem{corol}[defi]{Corollary}

\newtheorem{conj}[defi]{Conjecture}

\newtheorem{examp}[defi]{Example}

\theorembodyfont{\rmfamily}

\theorembodyfont{}
\newenvironment{prf}{\noindent\textbf{Proof:}\;}{\par\noindent\ignorespacesafterend}

\newcommand{\Qed}{\hfill $\square$}
\newcounter{dDef} 

\newcounter{dLem} 

\newcounter{dThm} 

\newcounter{dPro} 

\newcounter{Beispielzaehler}

\nc{\bm}{\boldmath}
\nc{\bmm}[1]{\mbox{\bm$\DST #1$}}
\nc{\mi}[1]{\bmm{\mathrm{(#1):}} \quad}
\usepackage[active]{srcltx}
\usepackage{a4}
\usepackage[all,poly]{xy}
\usepackage{float}

\usepackage{pifont}
\newcommand{\tick}{\ding{52}}
\usepackage{makecell}

\nc{\trighyp}[2]{T_{\hspace{-0.08ex}#2}(#1)}

\DMO{\thardness}{thd}

\DMO{\exstrahler}{HS}
\nc{\exhst}[2]{\exstrahler(#1,#2)}
\nc{\exstrahlersize}[2]{\alpha(#1,#2)}

\DMO{\smuo}{F^1}
\DMO{\tsmuo}{T^1}
\nc{\Tsmuo}{\mc{T}_1}

\nc{\EUrefc}{\exists\Urefc}
\nc{\UrefcE}{\Urefc[\exists]}

\DMO{\nosim}{\nrightarrow}
\DMO{\nosima}{\nrightarrow_a}

\nc{\Up}{\mc{UP}}

\begin{document}

\title{Trading inference effort versus size in\\ CNF Knowledge Compilation}

\author{Matthew Gwynne and Oliver Kullmann\\
  \qquad {\small \url{http://cs.swan.ac.uk/~csmg/}} \qquad
  {\small \url{http://cs.swan.ac.uk/~csoliver}}\\
  \href{http://www.swan.ac.uk/compsci/}{Computer Science Department}\\
  \href{http://www.swan.ac.uk/}{Swansea University}\\
  Swansea, UK
}

\maketitle

\begin{abstract}
  Knowledge Compilation (KC) studies compilation of boolean functions $f$ into some formalism $F$, which allows to answer all queries of a certain kind in polynomial time. Due to its relevance for SAT solving, we concentrate on the query type ``clausal entailment'' (CE), i.e., whether a clause $C$ follows from $f$ or not, and we consider subclasses of CNF, i.e., clause-sets $F \in \Cls$ with special properties (CNF itself is not suitable for CE queries unless P=NP). In this report we do not allow auxiliary variables (except of the Outlook), and thus $F$ needs to be equivalent to $f$.

  We consider the hierarchies $\Urefc_k \sse \Wrefc_k \subset \Cls$ ($k \in \NNZ$), which were introduced in \cite{GwynneKullmann2012SlurSOFSEM,GwynneKullmann2012SlurJ}, and where each level allows CE queries. The first two levels are well-known classes for KC, namely $\Urefc_0 = \Wrefc_0$ is the same as PI as studied in KC, that is, $f$ is represented by the set of all prime implicates, while $\Urefc_1 = \Wrefc_1$ is the same as $\Urefc$, the class of unit-refutation complete clause-sets introduced in \cite{Val1994UnitResolutionComplete}. We show that for each $k$ there are (sequences of) boolean functions with polysize representations in $\Urefc_{k+1}$, but with an exponential lower bound on representations in $\Wrefc_k$. Such a separation was previously only know for $k=0$. We also consider $\Propc \subset \Urefc$, the class of propagation-complete clause-sets introduced in \cite{DarwichePipatsrisawat2011ClauseLearnRes,BordeauxMarquesSilva2012KnowledgeCompilation}. Strengthening \cite{BBCGKV2013Propc}, we show that there are (sequences of) boolean functions with polysize representations in $\Urefc$, while there is an exponential lower bound for representations in $\Propc$. These separations are steps towards a general conjecture determining the representation power of the hierarchies $\Propc_k \subset \Urefc_k \sse \Wrefc_k$. The strong form of this conjecture also allows auxiliary variables, as discussed in depth in the Outlook.
\end{abstract}

\tableofcontents

\section{Introduction}
\label{sec:intro}

Boolean functions $f: \set{0,1}^n \ra \set{0,1}$ are fundamental objects of computer science, and many fields are concerned with their representation.\footnote{See \cite{CramaHammer2011BooleanFunctions} for the basic theory, \cite{CramaHammer2010BooleanModels} for an overview on their applications, and \cite{Jukna2012BooleanFunctionComplexity} for the complexity theory of their circuit representations.} In Knowledge Compilation (KC; see \cite{DarwicheMarquis2002KCmap} for a general overview), $f$ is given by some propositional formula (theory), and is to be compiled (off-line, that is, complex computations are possible here) into some $F$ belonging to some target language, such that a large number of queries of a certain kind can be answered efficiently (using $F$).

A natural target language is CNF (conjunctive normal forms), for which we write ``$F \in \Cls$'', where $\Cls$ is the class of all clause-sets, interpreted as CNFs. A basic subclass is PI, that is $F$ is the (precisely) the set of all prime implicates of some boolean function $f$ (this is a true normal form for $f$, since it is unique and identifies $f$). Now in general not all prime implicates are needed, if additional mechanisms are used to answer queries. This led to the introduction of the class $\Urefc$ of ``unit-refutation complete clause-sets'' in \cite{Val1994UnitResolutionComplete}, where the defining property of $F \in \Urefc$ is that if instantiation, that is, applying a partial assignment $\vp$ to $F$, resulting in the clause-set $\vp * F$, yields an unsatisfiable $\vp * F$, then this is detected by unit-clause propagation. Using $\rk_1: \Cls \ra \Cls$ for the process of unit-clause propagation, detection of unsatisfiability means $\bot \in \rk_1(\vp * F)$, where $\bot$ is the empty clause (while the defining property of IP is that in that case we already have $\bot \in \vp * F$).\footnote{Note that $F \models C$ holds iff $\vp_C *F$ is unsatisfiable, where $\vp_C$ sets all literals in $C$ to $0$. So by the definition of $\Urefc$, the query-type ``clausal entailment'' (CE) is directly handled by $\rk_1$.} It is shown in \cite{Val1994UnitResolutionComplete} that there are short clause-sets in $\Urefc$ with an exponential number of prime implicates (and so the equivalent representation in IP is very large).

The question was raised of the worst-case growth when compiling from an arbitrary CNF clause-set $F$ to some equivalent $F' \in \Urefc$. A first approach can be seen in \cite{BKNW2009CircuitComplexity}, where the authors provide examples of constraints with only super-polynomial size CNF-representations with certain consistency guarantees, even when allowing auxiliary variables; this has been developed further in \cite{GwynneKullmann2013GoodRepresentationsII} (see Subsection \ref{sec:conclstrict}). This shows a super-polynomial lower-bound on the worst-case growth, but no method or new (larger) target-class for knowledge-compilation. Another partial answer was given in \cite{BBCGKV2013Propc}, where clause-sets are given where every equivalent clause-set in $\Propc \subset \Urefc$ is of exponential size. Our main result now answers the question of worst-case growth from \cite{Val1994UnitResolutionComplete} in full generality with the hierarchy $\mr{PI} = \Urefc_0 \subset \Urefc_1 = \Urefc \subset \Urefc_2 \subset \dots$. Each level of $\Urefc_k$ is exponentially more expressive than the previous one, i.e., with possible \emph{exponential} blow-up when compiling from some $F \in \Urefc_{k+1}$ to equivalent $F' \in \Urefc_k$. So each level offers a new, larger class for knowledge compilation, at the expense of increased query time ($O(\ell(F) \cdot n(F)^{2k-2})$ for $\Urefc_k$). This separation, between $\Urefc_{k+1}$ and $\Urefc_k$ for arbitrary $k$ is more involved than the simple separation in \cite{Val1994UnitResolutionComplete}, due to the parameterised use of more advanced polynomial-time methods than $\rk_1$, while the separation between $\Urefc_0$ and $\Urefc_1$ is actually rather simple, since $\Urefc_0$ does not allow any form of compression. To explain the hierarchy $\Urefc_k$ and $\Propc$, we need to connect to SAT solving.

\subsection{Hierarchies for CNF Knowledge Compilation}
\label{sec:introhcnfkc}

A basic task of KC is to find shortest (or short) representations in the target class. This has also applications in the area of ``SAT solving'', which is about deciding satisfiability of propositional formulas, mostly in CNF.\footnote{See \cite{2008HandbuchSAT} for an overview.} Often the translation starts with a set of boolean constraints (in fact boolean functions), and size of the translation is a basic criterion to be optimised. Furthermore, the target class should be ``easy'' for SAT solving. The quest for such classes of clause-sets with polynomial-time SAT-decision led to the hierarchy $\Urefc_k \subset \Cls$, $k \in \NNZ$, with $\Urefc_k \subset \Urefc_{k+1}$ and $\bc_k \Urefc_k = \Cls$, where $\Urefc_0$ is PI, and $\Urefc_1 = \Urefc$, by the following development.\footnote{To be fully precise, $\Urefc_0$ is the class of clause-sets such that after elimination of subsumed clauses we obtain an element of PI.}

The class $\Slur$ (``Single Lookahead Unit Resolution'') was introduced in \cite{SAFS95} as an umbrella class for efficient SAT solving. \cite{CepekKuceraVlcek2012SLUR,BalyoGurskyKuceraVlcek2012SLURHier} extended this class in various ways to hierarchies covering all of CNF (all clause-sets). These hierarchies were unified and strengthened in \cite{GwynneKullmann2012SlurSOFSEM,GwynneKullmann2012SlurJ} to the classes $\Slur_k$, with $\Slur_1 = \Slur$, using generalised unit-clause propagation $\rk_k: \Cls \ra \Cls$ as introduced in \cite{Ku99b}. The well-known case of full failed-literal elimination is precisely $\rk_2$, which applies a reduction $F \leadsto \pao x1 * F$ as long as there is a literal $x$ with $\bot \in \rk_1(\pao x0 * F)$, and $\rk_k$ is the natural generalisation to arbitrary $k$. The class $\Slur_k$ is the class of all clause-sets $F$, where either $\bot \in \rk_k(F)$, or else one is guaranteed to find a satisfying assignment by choosing any literal $x$ with $\bot \notin \rk_k(\pao x1 * F)$, reducing $F \leadsto \pao x1 * F$, and repeating this process. Using $\rk_k$, we can also define $\Urefc_k$ as the class of clause-sets $F$ such that for each partial assignment $\vp$ with unsatisfiable $\vp * F$ holds $\bot \in \rk_k(\vp * F)$. A basic result of \cite{GwynneKullmann2012SlurSOFSEM,GwynneKullmann2012SlurJ} is that $\Slur_k = \Urefc_k$ holds for all $k \in \NNZ$, which further motivates the claim that representation of boolean functions via $\Urefc_k$ has special relevance for finding good SAT translations.

The basic hierarchy $\Urefc_k$ had two offsprings, the stricter hierarchy $\Propc_k$ and the wider hierarchy $\Wrefc_k$.
Generalising the class $\Propc \subset \Urefc$ of ``unit-propagation complete clause-sets'', introduced in  \cite{BordeauxMarquesSilva2012KnowledgeCompilation} (using ideas from \cite{DarwichePipatsrisawat2011ClauseLearnRes}), the interleaving hierarchy $\Propc_k$, with $\Propc_0 \subset \Urefc_0 \subset \Propc_1 \subset \Urefc_1 \subset \dots$ was defined in \cite{GwynneKullmann2012SlurJ}, and further studied in \cite{GwynneKullmann2013GoodRepresentationsII}. The elements of $\Propc_k$ are those clause-sets $F$ such that for each partial assignment $\vp$ either $\bot \in \rk_k(\vp * F)$ holds or otherwise $\rk_k(\vp * F)$ does not have any forced assignments. The hierarchy $\Wrefc_k$ with $\Wrefc_0 = \Urefc_0$, $\Wrefc_1 = \Urefc_1$ and $\Wrefc_k \supset \Urefc_k$ for $k \ge 2$, also defined in \cite{GwynneKullmann2012SlurJ}, and further studied in \cite{GwynneKullmann2013GoodRepresentationsII}, is defined as the class of clause-sets $F$ such that for each partial assignment $\vp$ with unsatisfiable $\vp * F$ the inconsistency of $\vp * F$ can be derived by $k$-resolution, that is, resolution where for each resolution step at least one parent clause has length at most $k$.

In this report we consider these hierarchies $\Propc_k, \Urefc_k, \Wrefc_k$ for the purpose of KC, representing boolean functions by equivalent clause-sets in one of these classes. Conjecture 1.1 in \cite{GwynneKullmann2012SlurJ} says that there are boolean functions with short equivalent clause-sets in $\Urefc_{k+1}$, but without short equivalent clause-sets in $\Urefc_k$, for each $k$. While Conjecture 9.9 in \cite{GwynneKullmann2012SlurJ} says, when considered for the case without auxiliary variables, the same for the hierarchy $\Wrefc_k$. We show both separations together, in a stronger form, in Theorem \ref{thm:separation}, namely we show that there are short clause-sets in $\Urefc_{k+1}$ which have no short equivalent clause-sets in $\Wrefc_k$. Furthermore we show that there are short clause-sets in $\Urefc$ without equivalent short clause-sets in $\Propc$.

\subsection{Mapping the hierarchies}
\label{sec:intromap}

Our separation results show parts of a general conjecture, which determines the relations between the classes of the three hierarchies $\Propc_k, \Urefc_k, \Wrefc_k$ regarding their expressive power w.r.t.\ equivalence. First we need some definitions:
\begin{itemize}
\item For a clause-set $F \in \Cls$ we use $\bmm{n(F)} := \var(F)$ for the number of variables and $\bmm{\ell(F)} := \sum_{C \in F} \abs{C}$ for the number of literal occurrences.
\item A sequence $(F_n')_{n \in \NN}$ of clause-sets is \textbf{equivalent} to a sequence of $(F_n)_{n \in \NN}$, if $F_n'$ is equivalent to $F_n$ for each $n \in \NN$.
\end{itemize}
Now we can define precisely what it means that a class $\mc{C}$ of clause-sets can be more succinct than another class $\mc{C}'$:
\begin{defi}\label{def:nosim}
  For $\mc{C}, \mc{C}' \sse \Cls$ the relation \bmm{\mc{C}' \nosim \mc{C}} (``$\mc{C}'$ does not simulate $\mc{C}$'') holds if there is a sequence $(F_n)_{n \in \NN}$ in $\mc{C}$ (i.e., $F_n \in \mc{C}$) such that $n(F_n) = n$ and $F_n$ is computable in time $n^{O(1)}$, and such that there is no equivalent sequence $(F_n')_{n \in \NN}$ in $\mc{C}'$ with $\ell(F_n') = n^{O(1)}$.\footnote{The condition on the number of variables restricts the boolean functions to some form of ``simple'' functions (which have a short representation in the number of variables). Sequences $(F_m)_{m \in \NN}$ with $n(F_m) = \Omega(m)$ are more convenient to handle, and are converted to standard form ``$n(F_m) = m$'' via appropriate forms of padding.}
\end{defi}
The main conjecture (weak form) now says, that the subset-relations between the classes we consider already determine their expressive power (while the strong form, Conjecture \ref{con:noauxstrstrf}, also allows the use of auxiliary variables, and is discussed in the conclusions):
\begin{conj}[Main Conjecture, weak form]\label{con:noauxstr}
  For $\mc{C}, \mc{C}' \in \set{\Propc_k, \Urefc_k, \Wrefc_k : k \in \NNZ}$ we have $\mc{C} \nosim \mc{C}'$ if and only if $\mc{C}' \not\sse \mc{C}$.
\end{conj}
If follows from Conjecture \ref{con:noauxstr} that for these classes $\mc{C}, \mc{C}'$ there is a polytime-computable map translating every clause-set in $\mc{C}$ into an equivalent clause-set in $\mc{C}'$ if and only if $\mc{C} \sse \mc{C}'$ (where the map is just the identity). The relation $\mc{C} \nosim \mc{C}'$ is stronger than $\mc{C} \not\le \mc{C}'$, where $\mc{C} \le \mc{C}'$ is the relation ``$\mc{C}$ is at least as succinct as $\mc{C}'$'' as defined for example in \cite{DarwicheMarquis2002KCmap}, since we do not require that a single polynomial regulates the size-relation between representations via these classes (as in $\mc{C} \le \mc{C}'$), but for every sequence there can be another polynomial (and moreover, we only consider ``simple sequences'').

Our main result, Theorem \ref{thm:separation}, is $\Wrefc_k \nosim \Urefc_{k+1}$ for every $k \in \NNZ$, that is, there are polysize sequences in $\Urefc_{k+1}$ such that no equivalent polysize sequences exist in $\Wrefc_k$ (moreover we show an exponential separation). We also show $\Propc \nosim \Urefc$ (Theorem \ref{thm:sepUCPC}; again, we show in fact an exponential separation). The remaining open cases of Conjecture \ref{con:noauxstr} are discussed in Subsection \ref{sec:concfsep}.

\subsection{Understanding the structure of satisfiable clause-sets}
\label{sec:introcombsat}

To be able to prove properties about all equivalent representations of some clause-set $F$, we must be able to understand its combinatorial structure in relation to the set of all its prime implicates. The notion of minimal unsatisfiability (MU) and minimally unsatisfiable subsets (MUS) is important in understanding the combinatorics of unsatisfiable clause-sets (see \cite{Kullmann2007HandbuchMU,MarquesSilva2012MUS}). To understand the structure of satisfiable clause-sets and their associated boolean functions, we now consider the concept of ``minimal premise sets'' (MPS) introduced in \cite{Kullmann2007ClausalFormZII}. The notion of MPS generalises that of MU by considering clause-sets $F$ which are minimal w.r.t implying \emph{any} clause $C$ rather than just those implying $\bot$. And accordingly we consider the minimal-premise subsets (MPSS) of a clause-set $F$.

Every prime implicate $C$ of a clause-set $F$ has an associated MPSS (just consider the minimal sub-clause-set of $F$ that implies $C$), but not every MPSS of $F$ yields a prime implicate (e.g., consider the MPSS $\set{C}$ for some non-prime clause $C \in F$). However, by ``doping'' the clause-set, i.e., adding a new unique variable to every clause, every clause in an MPSS $F'$ makes a unique contribution to its derived clause $C$. This results in a new clause-set $\doping(F)$ which has an exact correspondence between its minimal premise sets (which are (essentially) also those of $F$) and its prime implicates. In this way, by considering clause-sets $F$ with a very structured set of minimal premise subsets, we can derive clause-sets $\doping(F)$ with very structured set of prime implicates.

\subsection{Finding relatively hard boolean functions}
\label{sec:introstricteq}

A sequence $(f_h)_{h \in \NN}$ of boolean functions, which separates $\Urefc_{k+1}$ from $\Urefc_k$ w.r.t.\ clause-sets equivalent to $f_h$ in $\Urefc_{k+1}$ resp.\ $\Urefc_k$, should have the following properties:
\begin{enumerate}
\item \textbf{A large number of prime implicates}: the number of prime implicates for $f_h$ should at least grow super-polynomially in $h$, since otherwise already the set of prime implicates is a small clause-set in $\Urefc_0$ equivalent to $f_h$.
\item \textbf{Easily characterised prime implicates}: the prime implicates of $f_h$ should be easily characterised, since otherwise we can not understand how clause-sets equivalent to $f_h$ look like.
\item \textbf{Poly-size representations}: there must exist short clause-sets in $\Urefc_{k+1}$ equivalent to $f_h$ for all $h \in \NN$.
\end{enumerate}

\cite{SloanSzoerenyiTuran2005Primimplikanten_1} introduced a special type of boolean functions, called \ul{N}on-repeating \ul{U}nate \ul{D}ecision trees (NUD) there, by adding new variables to each clause of clause-sets in $\Smusati{\delta=1}$, which is the class of unsatisfiable hitting clause-sets of deficiency $\delta=1$. These boolean functions have a large number of prime implicates (the maximum regarding the original number of clauses), and thus are natural to consider as candidates to separate the levels of $\Urefc_k$. In Section \ref{sec:minpsdopingcls} we show that the underlying $\Smusat_{\delta=1}$ clause-sets determine the structure. The clause-sets in $\Smusat_{\delta=1}$ are exactly those with the maximum number of minimal premise sets, and then doping elements of $\Smusat_{\delta=1}$ yields clause-sets with the maximal number of prime implicates. We utilise the tree structure of $\Smusat_{\delta=1}$ to prove lower bounds on the size of equivalent representations in $\Urefc_k$ of doped $\Smusat_{\delta=1}$ clause-sets.

In Section \ref{sec:lowerb} we introduce the basic method (see Theorem \ref{thm:triggersetmethod}) for lower bounding the size of equivalent clause-sets of a given hardness, via the transversal number of ``trigger hypergraphs''. The basic idea is very simple, namely if we want $F$ to have hardness at most $k$, then for every prime implicate $C$ of $F$ the (unsatisfiable) clause-set $\vp_C * F$ must contain a clause of length at most $k$, in order to ``trigger'' the derivation of the empty clause from $\vp_C * F$. This applies to w-hardness as well, and thus we actually obtain a lower bound on the w-hardness.

Using this lower-bound method, in Theorem \ref{thm:nogoodksoft} we  show a lower bound on the matching number (the maximal number of disjoint hyperedges) of the trigger hypergraph of doped ``extremal'' $\Smusati{\delta=1}$-clause-sets. From this follows immediately Theorem \ref{thm:separation}, that for every $k \in \NNZ$ there are polysize clause-sets in $\Urefc_{k+1}$, where every equivalent clause-set in $\Wrefc_k$ is of exponential size. Thus the $\Urefc_k$ as well as the $\Wrefc_k$ hierarchy is strict regarding equivalence of polysize clause-sets.

\subsection{Relevance of these hierarchies for SAT solving}
\label{sec:introrelSAT}

The poly-time methods used to detect unsatisfiability of instantiations of clause-sets in $\Urefc_k$ resp.\ $\Wrefc_k$ have a running-time with an exponent depending on $k$, and in the latter case also space-complexity depends in the exponent on $k$.
\begin{enumerate}
\item This seems a necessary condition for showing a separation result as in this paper. It is needed that the different levels are qualitatively different. And this seems very unlikely to be achievable with a parameter which would allow fixed-parameter tractability, and which thus would only be a quantitative parameter (like the number of variables), only expressing a gradual increase in complexity.\footnote{Weaker means for deriving forced assignments than by $\rk_k$ have been considered in \cite{HaanKanjSzeider2013Backbones}.} See Lemma \ref{lem:collapsecanon} for an example of a collapsing hierarchy.
\item The class $\Urefc_k$ uses generalised unit-clause propagation, namely the reduction $\rk_k$. Especially $\rk_2$, which is (complete) failed-literal elimination, is used in look-ahead SAT solvers (see \cite{HvM09HBSAT} for an overview) such as \OKsolver{} (\cite{Ku2002h}), \texttt{march} (\cite{HeuleDufourvanZwietenMaaren2004Reasoning}) and \texttt{satz} (\cite{LA1996}). Also conflict-driven solvers such as \texttt{CryptoMiniSat} (\cite{Soos2010CMSDesc}) and \texttt{PicoSAT} (\cite{Biere2008picosat,Biere2010PicoSATLingelingDesc}) integrate $\rk_2$ during search, and solvers such as \texttt{Lingeling} (\cite{Biere2010PicoSATLingelingDesc,Biere2012LingelingDesc}) use $\rk_2$ as a preprocessing technique. Furthermore, in general $\rk_k$ is used, in even stronger versions, in the St\r{a}lmarck-solver (see \cite{SS90,Har96,SSt98}, and see Section 3.5 of \cite{Ku99b} for a discussion of the connections to $\rk_k$), and via breadth-first ``branch/merge'' rules in \texttt{HeerHugo} (see \cite{GW00}).
\end{enumerate}

\subsection{Overview on results}

The preliminaries (Section \ref{sec:prelim}) define the basic notions. The classes $\Urefc_k$, $\Propc_k$ and $\Wrefc_k$ are defined in Section \ref{sec:measurerepcomp}. In Section \ref{sec:minpsdopingcls} we investigate minimal premise sets and doping in general, while in Section \ref{sec:doptreecls} we apply these notions to our source of hard examples. In Section \ref{sec:lowerb} we are then able to show the separation $\Wrefc_k \nosim \Urefc_{k+1}$. In Section \ref{sec:hierkcprop} we discuss the KC-queries supported by our three hierarchies. In Section \ref{sec:sepUCPC} we show $\Propc \nosim \Urefc$. Finally, in Section \ref{sec:open} one finds many open problems. We now list our mean results (marked as ``theorems'', in contrast to ``lemmas'', which are ``small results''). The main results on minimal premise sets and doping are:
\begin{enumerate}
\item Theorem \ref{thm:dopedmpsprc} shows the correlation between prime implicates of doped clause-sets and minimal premise-sets of the original (undoped) clause-sets.
\item Theorem \ref{thm:dopedsmumax} characterises unsatisfiable clause-sets where every non-empty sub-clause-set is a minimal premise set.
\item Theorem \ref{thm:sumdsmuo} gives basic characteristics of doped $\Smusati{\delta=1}$-clause-sets.
\end{enumerate}

The main results related to the three hierarchies are:
\begin{enumerate}
\item Theorem \ref{thm:triggersetmethod} introduces the basic method for lower bounding the size of equivalent clause-sets of a given w-hardness, via the transversal number of ``trigger hypergraphs''.
\item Theorem \ref{thm:nogoodksoft} shows a lower bound on the matching number of the trigger hypergraph of doped ``extremal'' $\Smusati{\delta=1}$-clause-sets.
\item Theorem \ref{thm:separation} shows that for every $k \in \NNZ$ there are polysize clause-sets in $\Urefc_{k+1}$, where every equivalent clause-set in $\Wrefc_k$ is of exponential size.
\item Theorem \ref{thm:kcprop} states KC queries supported by the three hierarchies.
\item Theorem \ref{thm:sepUCPC} shows that there are polysize clause-sets in $\Urefc$, where every equivalent clause-set in $\Propc$ is of exponential size.
\end{enumerate}

\paragraph{Remarks on the history of this report}

Many results of this report were originally contained in \cite{GwynneKullmann2013GoodRepresentations}. That report, conceived as a starting point for a theory of SAT representations, had three topics: The separation results as in this paper, representation of XOR constraints, and the relations to SAT solving. The fifth version would have had over 80 pages, and so we decided to split it into three reports (which each contain substantial additions):
\begin{enumerate}
\item The representation of XOR constraints is now in \cite{GwynneKullmann2013GoodRepresentationsII}.
\item Results regarding the separation of the hierarchies in this report.
\item While the SAT-related theory and experimentation is in \cite{GwynneKullmann2013GoodRepresentationsIII} (to appear).
\end{enumerate}

\section{Preliminaries}
\label{sec:prelim}

We follow the general notations and definitions as outlined in \cite{Kullmann2007HandbuchMU}. We use $\NN = \set{1,2,\dots}$, $\NNZ = \NN \cup \set{0}$, and $\pot(M)$ for the set of subsets of set $M$.

\subsection{Clause-sets}
\label{sec:prelimcls}

Let $\Va$ be the infinite set of variables, and let $\Lit = \Va \cup \set{\ol{v} : v \in \Va}$ be the set of literals, the disjoint union of variables as positive literals and complemented variables as negative literals. We use $\ol{L} := \set{\ol{x} : x \in L}$ to complement a set $L$ of literals. A clause is a finite subset $C \subset \Lit$ which is complement-free, i.e., $C \cap \ol{C} = \es$; the set of all clauses is denoted by $\Cl$. A clause-set is a finite set of clauses, the set of all clause-sets is $\Cls$. By $\var(x) \in \Va$ we denote the underlying variable of a literal $x \in \Lit$, and we extend this via $\var(C) := \set{\var(x) : x \in C} \subset \Va$ for clauses $C$, and via $\var(F) := \bc_{C \in F} \var(C)$ for clause-sets $F$. The possible literals in a clause-set $F$ are denoted by $\lit(F) := \var(F) \cup \ol{\var(F)}$. Measuring clause-sets happens by $n(F) := \abs{\var(F)}$ for the number of variables, $c(F) := \abs{F}$ for the number of clauses, and $\ell(F) := \sum_{C \in F} \abs{C}$ for the number of literal occurrences. A special clause-set is $\top := \es \in \Cls$, the empty clause-set, and a special clause is $\bot := \es \in \Cl$, the empty clause.

A partial assignment is a map $\vp: V \ra \set{0,1}$ for some finite $V \subset \Va$, where we set $\var(\vp) := V$, and where the set of all partial assignments is $\Pass$. For $v \in \var(\vp)$ let $\vp(\ol{v}) := \ol{\vp(v)}$ (with $\ol{0} = 1$ and $\ol{1} = 0$). We construct partial assignments by terms $\pab{x_1 \ra \ve_1, \dots, x_n \ra \ve_n} \in \Pass$ for literals $x_1, \dots, x_n$ with different underlying variables and $\ve_i \in \set{0,1}$. We use $\vp_C := \pab{x \ra 0 : x \in C}$ for the partial assignment setting precisely the literals in clause $C \in \Cl$ to false.

For $\vp \in \Pass$ and $F \in \Cls$ we denote the result of applying $\vp$ to $F$ by $\vp * F$, removing clauses $C \in F$ containing $x \in C$ with $\vp(x) = 1$, and removing literals $x$ with $\vp(x) = 0$ from the remaining clauses. By $\Sat := \set{F \in \Cls \mb \ex\, \vp \in \Pass : \vp * F = \top}$ the set of satisfiable clause-sets is denoted, and by $\Usat := \Cls \sm \Sat$ the set of unsatisfiable clause-sets.

So clausal entailment, that is the relation $F \models C$ for $F \in \Cls$ and $C \in \Cl$, which by definition holds true iff for all $\vp \in \Pass$ with $\vp * F = \top$ we have $\vp * \set{C} = \top$, is equivalent to $\vp_C * F \in \Usat$.

Two clauses $C, D \in \Cl$ are resolvable iff they clash in exactly one literal $x$, that is, $C \cap \ol{D} = \set{x}$, in which case their resolvent is $\bmm{C \res D} := (C \cup D) \sm \set{x,\ol{x}}$ (with resolution literal $x$). A resolution tree is a full binary tree formed by the resolution operation. We write \bmm{T : F \vdash C} if $T$ is a resolution tree with axioms (the clauses at the leaves) all in $F$ and with derived clause (at the root) $C$. A resolution tree $T : F \vdash C$ is regular iff along each path from the root of $T$ to a leaf no resolution-variable is used more than once. In this article we use only resolution \emph{trees}, even when speaking of unrestricted resolution, that is, we always unfold dag-resolution proofs to (full) binary resolution trees. Completeness of resolution means that $F \models C$ (semantic implication) is equivalent to $F \vdash C$, i.e., there is some $C' \sse C$ and some $T$ with $T: F \vdash C'$.

A \emph{prime implicate} of $F \in \Cls$ is a clause $C$ such that a resolution tree $T$ with $T: F \vdash C$ exists, but no $T'$ exists for some $C' \subset C$ with $T': F \vdash C'$; the set of all prime implicates of $F$ is denoted by $\bmm{\primec_0(F)} \in \Cls$. The term ``implicate'' refers to the implicit interpretation of $F$ as a conjunctive normal form (CNF). Considering clauses as combinatorial objects one can speak of ``prime clauses'', and the ``$0$'' in our notation reminds of ``unsatisfiability'', which is characteristic for CNF. Two clause-sets $F, F' \in \Cls$ are equivalent iff $\primec_0(F) = \primec_0(F')$. A clause-set $F$ is unsatisfiable iff $\primec_0(F) = \set{\bot}$. The set of \emph{prime implicants} of a clause-set $F \in \Cls$ is denoted by $\bmm{\primec_1(F)} \in \Cls$, and is the set of all clauses $C \in \Cl$ such that for all $D \in F$ we have $C \cap D \not= \es$, while this holds for no strict subset of $C$.

\subsection{On ``good'' equivalent clause-sets}
\label{sec:prelimgoodeqcls}

A basic problem considered in this article is for a given $F \in \Cls$ to find a ``good'' equivalent $F' \in \Cls$. How ``good'' $F'$ is depends in our context on two factors, which have to be balanced against each other:
\begin{itemize}
\item the size of $F'$: we measure $c(F')$, and the smaller the better;
\item the inference power of $F'$: inference from $F'$ should be ``as easy as possible'', and we consider two measures in this article, (tree-)hardness in Subsection \ref{sec:prelimhdUC}, and width-hardness in Subsection \ref{sec:prelimwhdWC}; the smaller these measures, the easier inference w.r.t.\ tree resolution resp.\ (generalised) width-bounded resolution.
\end{itemize}
The basic size-lower-bound for $F'$ is given by the \textbf{essential prime implicates}, which are those $C \in \primec_0(F)$ such that $\primec_0(F) \sm \set{C}$ is not equivalent to $F$:
\begin{lem}\label{lem:necprcls}
  Consider $F \in \Cls$, and let $P \sse \primec_0(F)$ be the set of essential prime implicates of $F$. Now for every $F' \in \Cls$ equivalent to $F$ there exists an injection $i: P \ra F'$ such that for all $C \in P$ holds $C \sse i(C)$. Thus $c(F') \ge c(P)$.
\end{lem}
\begin{prf}
  For every $C' \in F'$ there exists a $C \in \primec_0(F)$ such that $C \sse C'$; replacing every $C' \in F$ by such a chosen $C$ we obtain $F'' \sse \primec_0(F)$ with $P \sse F''$. \Qed
\end{prf}

Note that Lemma \ref{lem:necprcls} crucially depends on not allowing auxiliary variables --- when allowing new variable, then we currently do not have any overview on the possibilities for ``better'' $F'$. The most powerful representation regarding inference alone (with or without new variables) is given by the set $\primec_0(F)$ of all prime implicates of $F$, and will have ``hardness'' $0$, as defined in the following section. (The problem is of course that in most cases this representation is too large, and thus higher hardness must be allowed.)

\section{Measuring ``SAT representation complexity''}
\label{sec:measurerepcomp}

In this section we define and discuss the measures $\hardness, \phardness, \whardness: \Cls \ra \NNZ$ and the corresponding classes $\Urefc_k, \Propc_k, \Wrefc_k \subset \Cls$. It is mostly of an expository nature, explaining what we need from \cite{Ku99b,Ku00g,GwynneKullmann2012SlurSOFSEM,GwynneKullmann2012Slur,GwynneKullmann2012SlurJ}, with some additional remarks.

\subsection{Hardness and $\Urefc_k$}
\label{sec:prelimhdUC}

First we turn to the most basic hardness measurement. It can be based on resolution refutation trees, as we do here, but it can also be defined algorithmically, via generalised unit-clause propagation (see Lemma \ref{lem:charachd}).
\begin{defi}\label{def:hthts}
  For a full binary tree $T$ the height $\bmm{\height(T)} \in \NNZ$ and the Horton-Strahler number $\bmm{\hts(T)} \in \NNZ$ are defined as follows:
  \begin{enumerate}
  \item If $T$ is trivial (i.e., $\nnds(T) = 1$), then $\height(T) := 0$ and $\hts(T) := 0$.
  \item Otherwise let $T_1, T_2$ be the two subtrees of $T$:
    \begin{enumerate}
    \item $\height(T) := 1 + \max(\height(T_1), \height(T_2))$
    \item If $\hts(T_1) = \hts(T_2)$, then $\hts(T) := 1 + \max(\hts(T_1), \hts(T_2))$, otherwise $\hts(T) := \max(\hts(T_1), \hts(T_2))$.
    \end{enumerate}
  \end{enumerate}
\end{defi}
Obviously we always have $\hts(T) \le \height(T)$.
\begin{examp}\label{exp:hts}
  For the tree $T$ from Example \ref{exp:dopedsmutree} we have $\height(T) = 3$, $\hts(T) = 2$. The Horton-Strahler numbers of the subtrees are as follows:
    \begin{displaymath}
    \xygraph{
      !{0;/r8ex/:}
        []{2} (
          - [dll]{2} (
            -[dll]{1} (
              -[dl]{0} (),
              -[dr]{0} ()
            ),
            -[drr]{1} (
              -[dl]{0} (),
              -[dr]{0} ()
            )
          ),
          - [drr]{1} (
            -[dl]{0} (),
            -[dr]{0} ()
          )
        )
    }
  \end{displaymath}
\end{examp}

\begin{defi}\label{def:hardness}
  The hardness $\hardness: \Cls \ra \NNZ$ is defined for $F \in \Cls$ as follows:
  \begin{enumerate}
  \item If $F \in \Usat$, then $\hardness(F)$ is the minimum $\hts(T)$ for $T : F \vdash \bot$.
  \item If $F = \top$, then $\hardness(F) := 0$.
  \item If $F \in \Sat \sm \set{\top}$, then $\hardness(F) := \max_{\vp \in \Pass} \set{\hardness(\vp * F) : \vp * F \in \Usat}$.
  \end{enumerate}
\end{defi}

Hardness for unsatisfiable clause-sets was introduced in \cite{Ku99b,Ku00g}, while this generalisation to arbitrary clause-sets was first mentioned in \cite{AnsoteguiBonetLevyManya2008Hardness}, and systematically studied in \cite{GwynneKullmann2012SlurSOFSEM,GwynneKullmann2012Slur,GwynneKullmann2012SlurJ}. It is easy to see that the hardness of $F \in \Cls$ is the minimal $k \in \NNZ$ such that for all prime implicates $C$ of $F$ there exists $T : F \vdash C$ with $\hts(T) \le k$.

Definition \ref{def:hardness} defines hardness proof-theoretically; importantly, it can also be characterised algorithmically via necessary levels of generalised unit-clause propagation (see \cite{GwynneKullmann2012SlurSOFSEM,GwynneKullmann2012Slur,GwynneKullmann2012SlurJ} for the details):
\begin{lem}\label{lem:charachd}
  Consider the reductions $\rk_k: \Cls \ra \Cls$ for $k \in \NNZ$ as introduced in \cite{Ku99b}; it is $\rk_1$ unit-clause propagation, while $\rk_2$ is (full, iterated) failed-literal elimination. Then $\hardness(F)$ for $F \in \Cls$ is the minimal $k \in \NNZ$ such that for all $\vp \in \Pass$ with $\vp * F \in \Usat$ holds $\rk_k(\vp * F) = \set{\bot}$, i.e., the minimal $k$ such that $\rk_k$ detects unsatisfiability of any instantiation.
\end{lem}
For $F \in \Cls$ there is a partial assignment $\vp$ with $\vp * F = \rk_k(F)$, where $\vp$ consists of certain ``forced assignments'' $\pao x1 \sse \vp$, i.e., $\pao x0 * F \in \Usat$. Another ``localisation'' of forced assignments has been considered in \cite{HaanKanjSzeider2013Backbones}, namely ``$k$-backbones'', which is a forced assignment $\pao x1$ for $F$ such that there is $F' \sse F$ with $c(F') \le k$ and such that $\pao x1$ is forced also for $F'$. It is not hard to see that $\rk_k$ for $k \in \NNZ$ will set all $k$-backbones of $F \in \Cls$ (using that for $F \in \Usat$ we have $\hardness(F) < c(F)$ by Lemma 3.18 in \cite{Ku99b}).

We can now define our main hierarchy, the $\Urefc_k$-hierarchy (with ``UC'' for ``unit-refutation complete'') via (tree-)hardness:
\begin{defi}\label{def:UC}
  For $k \in \NNZ$ let $\bmm{\Urefc_k} := \set{F \in \Cls : \hardness(F) \le k}$.
\end{defi}
$\Urefc_1 = \Urefc$ is the class of unit-refutation complete clause-sets, as introduced in \cite{Val1994UnitResolutionComplete}. In \cite{GwynneKullmann2012SlurSOFSEM,GwynneKullmann2012Slur,GwynneKullmann2012SlurJ} we show that $\Urefc = \Slur$, where $\Slur$ is the class of clause-sets solvable via Single Lookahead Unit Resolution (see \cite{FrGe98}). Using \cite{CepekKuceraVlcek2012SLUR} we then obtain (\cite{GwynneKullmann2012SlurSOFSEM,GwynneKullmann2012Slur,GwynneKullmann2012SlurJ}) that membership decision for $\Urefc_k$ ($ = \Slur_k$) is coNP-complete for $k \ge 1$. The class $\Urefc_2$ is the class of all clause-sets where unsatisfiability for any partial assignment is detected by failed-literal reduction (see Section 5.2.1 in \cite{HvM09HBSAT} for the usage of failed literals in SAT solvers).

A basic fact is that the classes $\Urefc_k$ are stable under application of partial assignments, in other words, for $F \in \Cls$ and $\vp \in \Pass$ we have $\hardness(\vp * F) \le \hardness(F)$. For showing lower bounds on the hardness for unsatisfiable clause-sets, we can use the methodology developed in Subsection 3.4.2 of \cite{Ku99b}. A simplified version of Lemma 3.17 from \cite{Ku99b}, sufficient for our purposes, is as follows (with a technical correction, as explained in Example \ref{exp:lbhd}):
\begin{lem}\label{lem:lbhd}
  Consider $\mc{C} \sse \Usat$ and a function $h: \mc{C} \ra \NNZ$. For $k \in \NNZ$ let $\mc{C}_k := \set{F \in \mc{C} : h(F) \ge k}$. Then $\fa\, F \in \mc{C} : \hardness(F) \ge h(F)$ holds if and only if $\Urefc_0 \cap \mc{C}_1 = \es$, and for all $k \in \NN$, $F \in \mc{C}_k$ and $x \in \lit(F)$ there exist clause-sets $F_0, F_1 \in \Cls$ fulfilling the following three conditions:
  \begin{enumerate}
  \item[(i)] $n(F_{\ve}) < n(F)$ for both $\ve \in \set{0,1}$;
  \item[(ii)] $\hardness(F_{\ve}) \le \hardness(\pao x{\ve} * F)$ for both $\ve \in \set{0,1}$;
  \item[(iii)] $F_0 \in \mc{C}_k$ or $F_1 \in \mc{C}_{k-1}$.
  \end{enumerate}
\end{lem}
\begin{prf}
The given conditions are necessary for $\fa\, F \in \mc{C} : \hardness(F) \ge h(F)$, since we can choose $F_{\ve} := \pao v{\ve} * F$ for $\ve \in \set{0,1}$. To see sufficiency, assume for the sake of contradiction that there is $F \in \mc{C}$ with $\hardness(F) < h(F)$, and consider such an $F$ with minimal $n(F)$. If $\hardness(F) = 0$, so $h(F) = 0$ by assumption, and thus $\hardness(F) \ge 1$ would hold. So assume $\hardness(F) \ge 1$. It follows that there is a literal $x \in \lit(F)$ with $\hardness(\pao x1 * F) < \hardness(F)$. Let $k := h(F)$; so $F \in \mc{C}_k$. By assumption there are $F_0, F_1 \in \Cls$ with $\hardness(F_{\ve}) \le \hardness(\pao x{\ve} * F)$ for both $\ve \in \set{0,1}$, and $F_0 \in \mc{C}_k$ or $F_1 \in \mc{C}_{k-1}$. If $F_0 \in \mc{C}_k$, then $\hardness(F_0) \le \hardness(F) < k \le h(F_0)$, while $n(F_0) < n(F)$, contradicting minimality of $F$. And if $F_1 \in \mc{C}_{k-1}$, then $\hardness(F_1) \le \hardness(F) - 1 < k -1 \le h(F_1)$, while $n(F_1) < n(F)$, contradicting again minimality of $F$. \Qed
\end{prf}

Lemma 3.17 in \cite{Ku99b} doesn't state the condition (i) from Lemma \ref{lem:lbhd}. The following example shows that this condition actually needs to be stated (that is, if we just have (ii) and (iii), then $h$ doesn't need to be a lower bound for $\hardness$); fortunately in all applications in \cite{Ku99b} this (natural) condition is fulfilled.
\begin{examp}\label{exp:lbhd}
  Consider $\mc{C} := \Urefc_1 \cap \Usat$. Define $h: \mc{C} \ra \set{0,1,2}$ as $h(F) = 0$ iff $\bot \in F$, and $h(F) = 1$ iff $\bot \notin F$ and there is $v \in \var(F)$ with $\set{v},\set{\ol{v}} \in F$. So we have $h(F) = 2$ if and only if for all literals $x \in \lit(F)$ holds $\hardness(\pao x1 * F) = \hardness(\pao x0 * F) = 1$. By definition we have $\Urefc_0 \cap \mc{C}_1 = \es$. Now consider $k \in \set{1,2}$, $F \in \mc{C}_k$ and $x \in \lit(F)$. If $h(F) = 1$, then let $F_{\ve} := \pao x{\ve} * F$, while otherwise $F_{\ve} := F$ for $\ve \in \set{0,1}$. Now Conditions (ii), (iii) of Lemma \ref{lem:lbhd} are fulfilled (if $h(F) = 1$, then for Condition (iii) always $F_1 \in \mc{C}_{k-1}$ holds, while in case of $h(F) = 2$ we always have $F_0 \in \mc{C}_k$). But by definition $h$ is not a lower bound on $\hardness$.
\end{examp}

\subsection{P-Hardness and $\Propc_k$}
\label{sec:propc}

Complementary to ``unit-refutation completeness'', there is the notion of ``pro\-pa\-ga\-tion-com\-ple\-te\-ness'' as investigated in \cite{DarwichePipatsrisawat2011ClauseLearnRes,BordeauxMarquesSilva2012KnowledgeCompilation}, yielding the class $\Propc \subset \Urefc$. This was captured and generalised by a measure $\phardness: \Cls \ra \NNZ$ of ``propagation-hardness'' along with the associated hierarchy, defined in \cite{GwynneKullmann2012Slur,GwynneKullmann2012SlurJ} as follows:
\begin{defi}\label{def:phardness}
  For $F \in \Cls$ we define the \textbf{propagation-hardness} (for short ``p-hardness'') $\bmm{\phardness(F)} \in \NNZ$ as the minimal $k \in \NNZ$ such that for all partial assignments $\vp \in \Pass$ we have $\rk_k(\vp * F) = \rki(\vp * F)$, where $\rk_k: \Cls \ra \Cls$ is generalised unit-clause propagation (\cite{Ku99b,Ku00g}), and $\rki: \Cls \ra \Cls$ applies all forced assignments, and can be defined by $\rki(F) := \rk_{n(F)}(F)$. For $k \in \NNZ$ let $\bmm{\Propc_k} := \set{F \in \Cls : \phardness(F) \le k}$ (the class of \textbf{propagation-complete clause-sets of level $k$}).
\end{defi}
Remarks:
\begin{enumerate}
\item We have $\Propc = \Propc_1$.
\item For $k \in \NNZ$ we have $\Propc_k \subset \Urefc_k \subset \Propc_{k+1}$.
\item By definition (and composition of partial assignments) we have that all classes $\Propc_k$ are stable under application of partial assignments.
\item For $F \in \Cls$ a literal $x \in \Lit$ is \emph{forced for $F$} (more precisely, the assignment $\pao x1$ is forced for $F$), iff $\pao x0 * F \in \Usat$. Note that for $F \in\ \Usat$ all $x \in \Lit$ are forced, while for $F \in \Sat$ and a forced literal $x$ we have $x \in \lit(F)$. Now for $k \in \NNZ$ and $F \in \Cls$ we have $F \in \Propc_k$ iff for all $\vp \in \Pass$ the clause-set $F' := \rk_k(\vp * F)$ has no forced literals $x$ with $x \in \lit(F')$.
\end{enumerate}

\subsection{W-Hardness and $\Wrefc_k$}
\label{sec:prelimwhdWC}

A basic weakness of the standard notion of width-restricted resolution, which demands that \emph{both} parent clauses must have length at most $k$ for some fixed $k \in \NNZ$ (``width'', denoted by $\wid(F)$ below; see \cite{SW98}), is that even Horn clause-sets require unbounded width in this sense. The correct solution, as investigated and discussed in \cite{Ku99b,Ku00g}, is to use the notion of ``$k$-resolution'' as introduced in \cite{Kl93}, where only \emph{one} parent clause needs to have length at most $k$ (thus properly generalising unit-resolution). Nested input-resolution (\cite{Ku99b,Ku00g}) is the proof-theoretic basis of hardness, and approximates tree-resolution. In the same vein, $k$-resolution is the proof-theoretic basis of ``w-hardness'', and approximates dag-resolution (see Theorem 6.12 in \cite{Ku00g}):
\begin{defi}\label{def:whd}
  The \textbf{w-hardness} $\whardness: \Cls \ra \NNZ$ (``width-hardness'', or ``asymmetric width'') is defined for $F \in \Cls$ as follows:
  \begin{enumerate}
  \item If $F \in \Usat$, then $\whardness(F)$ is the minimum $k \in \NNZ$ such that $k$-resolution refutes $F$, that is, such that $T : F \vdash \bot$ exists where for each resolution step $R = C \res D$ in $T$ we have $\abs{C} \le k$ or $\abs{D} \le k$ (this corresponds to Definition 8.2 in \cite{Ku99b}, and is a special case of $\mr{wid}_{\mc{U}}$ introduced in Subsection 6.1 of \cite{Ku00g}).
  \item If $F = \top$, then $\whardness(F) := 0$.
  \item If $F \in \Sat \sm \set{\top}$, then $\DST \whardness(F) := \max_{\vp \in \Pass} \set{\whardness(\vp * F) : \vp * F \in \Usat}$.
  \end{enumerate}
  For $k \in \NNZ$ let $\bmm{\Wrefc_k} := \set{F \in \Cls : \whardness(F) \le k}$.

  The \textbf{symmetric width} $\wid: \Cls \ra \NNZ$ is defined in the same way, only that for $F \in \Usat$ we define $\wid(F)$ as the minimal $k \in \NNZ$ such that there is $T : F \vdash \bot$, where all clauses of $T$ (axioms and resolvents) have length at most $k$.
\end{defi}
Remarks:
\begin{enumerate}
\item We have $\Wrefc_0 = \Urefc_0$, $\Wrefc_1 = \Urefc_1$, and for all $k \in \NNZ$ holds $\Urefc_k \sse \Wrefc_k$ (this follows by Lemma 6.8 in \cite{Ku00g} for unsatisfiable clause-sets, which extends to satisfiable clause-sets by definition).
\item For $k \ge 3$ and $k' \ge 0$ we have $\Wrefc_k \cap \Usat \not\sse \Urefc_{k'}$; this follows from known resolution lower bounds for the symmetric width, for example in Subsection 10.2 of \cite{GwynneKullmann2013GoodRepresentationsII} a sequence $T_n$ of (short) unsatisfiable clause-sets with $\wid(T_n) = 3$ and $\hardness(T_n) = n$ is given.
\item Thus for $k \ge 3$ we have $\Urefc_k \subset \Wrefc_k$; Example \ref{exp:diffw2u2} extends this to $k \ge 2$.
\item Obviously we have $\whardness(F) \le \wid(F)$ for all $F \in \Cls$, where for $F \in \Ho \cap \Usat$ the symmetric width $\wid(F)$ is unbounded (actually it is precisely equal to the maximal clause-length of $F$), in contrast to $\whardness(F) \le 1$.
\end{enumerate}

\begin{examp}\label{exp:diffw2u2}
  An example for $F \in \Usat$ with $\whardness(F) = 2$ and $\hardness(F) = 3$ is
  \begin{multline*}
    F := \set{\set{2,3,4},\set{-4,2}, \; \set{-2,1,5},\set{-5,-2}, \; \set{-3,1,6},\set{-6,-3},\\
      \set{7,8,9},\set{-9,7}, \; \set{-7,-1,10},\set{-10,-7}, \; \set{-8,-1,11},\set{-11,-8}}.
  \end{multline*}
\end{examp}

We believe that this example can be extended:
\begin{conj}\label{con:diffw2u2}
  For $k \in \NNZ$ holds $\Wrefc_2 \not\sse \Urefc_k$.
\end{conj}

For unsatisfiable $F$, whether $\whardness(F) = k$ holds for $k \in \set{0,1,2}$ can be decided in polynomial time; this is non-trivial for $k = 2$ (\cite{BuroKleineBuening1996ResolutionShortClauses}) and unknown for $k > 2$. Nevertheless, the clausal entailment problem $F \models C$ for $F \in \Wrefc_k$ and fixed $k \in \NNZ$ is decidable in polynomial time, as shown in Subsection 6.5 of \cite{Ku00g}, by actually using a slight strengthening of $k$-resolution, which combines width-bounded resolution and input resolution. While space-complexity of the decision $F \models C$ for $F \in \Urefc_k$ is linear (for fixed $k$), now for $\Wrefc_k$ space-complexity is $O(\ell(F) \cdot n(F)^{O(k)})$.

As a special case of Theorem 6.12 in \cite{Ku00g} we obtain for $F \in \Usat$, $n(F) \not= 0$, the following general lower bound on resolution complexity:
\begin{displaymath}
  \compr(F) > b^{\frac{\whardness(F)^2}{n(F)}},
\end{displaymath}
where $b := e^{\frac 18} = 1.1331484 \ldots$, while $\compr(F) \in \NN$ is the minimal number of different clauses in a (tree-)resolution refutation of $F$. Similar to Theorem 14 in \cite{GwynneKullmann2012SlurSOFSEM} resp.\ Theorem 5.7 in \cite{GwynneKullmann2012Slur,GwynneKullmann2012SlurJ} we thus obtain:
\begin{lem}\label{lem:suffcritwhd}
  For $F \in \Cls$ and $k \in \NNZ$, such that for every $C \in \primec_0(F)$ with $\abs{C} < n(F)$ there exists a resolution proof of $C$ from $F$ using at most $b^{\frac{(k+1)^2}{n(F)-\abs{C}}}$ different clauses, we have $\whardness(F) \le k$.
\end{lem}

\section{Minimal premise sets and doped clause-sets}
\label{sec:minpsdopingcls}

In this section we study ``minimal premise sets'', ``mps's'' for short, introduced in \cite{Kullmann2007ClausalFormZII}, together with the properties of ``doped'' clause-sets, generalising a construction used in \cite{SloanSzoerenyiTuran2005Primimplikanten_1}. Mps's are generalisations of minimally unsatisfiable clause-sets stronger than irredundant clause-sets, while doping relates prime implicates and sub-mps's.

Recall that a clause-set $F$ is minimally unsatisfiable if $F \in \Usat$, while for all $C \in F$ holds $F \sm \set{C} \in \Sat$. The set of all minimally unsatisfiable clause-sets is $\bmm{\Musat} \subset \Cls$; see \cite{Kullmann2007HandbuchMU} for more information. In other words, for $F \in \Cls$ we have $F \in \Musat$ if and only if $F \models \bot$ and $F$ is minimal regarding this entailment relation. Now an mps is a clause-set $F$ which minimally implies some clause $C$, i.e., $F \models C$, while $F' \not\models C$ for all $F' \subset F$. In Subsection \ref{sec:minimalprems} we study the basic properties of mps's $F$, and determine the unique minimal clause implied by $F$ as $\purec(F)$, the set of pure literals of $F$.

For a clause-set $F$ its doped version $\doping(F) \in \Cls$ receives an additional new (``doping'') variable for each clause. The basic properties are studied in Subsection \ref{sec:Dopingcls}, and in Theorem \ref{thm:dopedmpsprc} we show that the prime implicates of $\doping(F)$ correspond 1-1 to the mps's contained in $F$. In Subsection \ref{sec:hddopedcls} we determine the hardness of doped clause-sets.

\subsection{Minimal premise sets}
\label{sec:minimalprems}

In Section 4.1 in \cite{Kullmann2007ClausalFormZII} basic properties of \emph{minimal premise sets} are considered:
\begin{defi}\label{def:Mps}
  A clause-set $F \in \Cls$ is a \textbf{minimal premise set} (``mps'') \textbf{for a clause $C \in \Cl$} if $F \models C$ and $\fa\, F' \subset F : F' \not\models C$, while $F$ is a \textbf{minimal premise set} if there exists a clause $C$ such that $F$ is a minimal premise set for $C$. The set of all minimal premise (clause-)sets is denoted by \bmm{\Mps}.
\end{defi}
Remarks:
\begin{enumerate}
\item $\top$ is not an mps (since no clause follows from $\top$).
\item An unsatisfiable clause-set is an mps iff it is minimally unsatisfiable, i.e., $\Mps \cap \Usat = \Musat$. In Corollary \ref{cor:mpswp} we will see that the minimally unsatisfiable clause-sets are precisely the mps's without pure literals.
\item Every minimal premise clause-set is irredundant (no clause follows from the other clauses).
\item For a clause-set $F$ and any implicate $F \models C$ there exists a minimal premise sub-clause-set $F' \sse F$ for C.
\item A single clause $C$ yields an mps $\set{C}$.
\item Two clauses $C \not= D$ yield an mps $\set{C,D}$ iff $C, D$ are resolvable.
\item If $F_1, F_2 \in \Mps$ with $\var(F_1) \cap \var(F_2) = \es$, then $F_1 \cup F_2 \notin \Mps$ except in case of $F_1 = F_2 = \set{\bot}$.
\end{enumerate}

\begin{examp}\label{exp:Mps}
  $\set{\set{a},\set{b}}$ for variables $a \not= b$ is irredundant but not an mps.
\end{examp}

With Corollary 4.5 in \cite{Kullmann2007ClausalFormZII} we see that no clause-set can minimally entail more than one clause:
\begin{lem}\label{lem:purelmps}
  For $F \in \Mps$ there exists exactly one $C \in \primec_0(F)$ such that $C$ is a minimal premise set for $C$, and $C$ is the smallest element of the set of clauses for which $F$ is a minimal premise set.
\end{lem}

We remark that Lemma \ref{lem:purelmps} does not mean that $\abs{\primec_0(F)} = 1$ for $F \in \Mps$; indeed, $F$ can have many $F' \subset F$ with $F' \in \Mps$, and each such $F'$ might contribute a prime implicate, as we will see later. We wish now to determine that unique prime implicate $C$ which follows minimally from an mps $F$. It is clear that $C$ must contain all pure literals from $F$, since all clauses of $F$ must be used, and we can not get rid off pure literals.

\begin{defi}\label{def:epc}
  For $F \in \Cls$ the \textbf{pure clause of $F$}, denoted by $\bmm{\purec(F)} \in \Cl$, is the set of pure literals of $F$, that is, $\purec(F) := L \sm (L \cap \ol{L})$, where $L := \bc F$ is the set of literals occurring in $F$.
\end{defi}

\begin{examp}\label{exp:epc}
  For $F = \set{\set{a,b},{\set{\ol{a},\ol{c}}}}$ we have $\purec(F) = \set{b,\ol{c}}$.
\end{examp}

The main observation for determining $C$ is that the conclusion of a regular resolution proof consists precisely of the pure literals of the axioms (this follows by definition):
\begin{lem}\label{lem:regrespurec}
  For a regular resolution proof $T: F \vdash C$, where every clause of $F$ is used as an axiom in $T$, we have $C = \purec(F)$.
\end{lem}

Due to the completeness of regular resolution we thus see, that $\purec(F)$ is the desired unique prime implicate:
\begin{lem}\label{lem:uniquepurec}
   For $F \in \Mps$ the unique prime implicate $C$, for which $F$ is a minimal premise set (see Lemma \ref{lem:purelmps}), is $C = \purec(F)$.
\end{lem}
\begin{prf}
  Consider a regular resolution proof $T: F \vdash C$ (recall that regular resolution is complete); due to $F \in \Mps$ every clause of $F$ must be used in $T$, and thus the assertion follows by Lemma \ref{lem:regrespurec}. \Qed
\end{prf}

\begin{corol}\label{cor:mpswp}
  If we have $F \in \Mps$ with $\purec(F) = \bot$, then $F \in \Musat$.
\end{corol}

By Lemma 4.4 in \cite{Kullmann2007ClausalFormZII} we get the main characterisation of mps's, namely that after elimination of pure literals they must be minimally unsatisfiable:
\begin{lem}\label{lem:characmps}
  Consider a clause-set $F \in \Cls$. Then $F \in \Mps$ if and only if the following two conditions hold for $\vp := \vp_{\purec(F)}$ (setting precisely the pure literals of $F$ to false):
  \begin{enumerate}
  \item $\vp * F \in \Musat$ (after removing the pure literals we obtain a minimal unsatisfiable clause-sets).
  \item $\vp$ is contraction-free for $F$, that is, for clauses $C, D \in F$ with $C \not= D$ we have $\vp * \set{C} \ne \vp * \set{D}$.
  \end{enumerate}
  These two conditions are equivalent to stating that $\vp * F$ as a multi-clause-set (not contracting equal clauses) is minimally unsatisfiable.
\end{lem}

Thus we obtain all mps's by considering some minimally unsatisfiable clause-sets and adding new variables in the form of pure literals:
\begin{corol}\label{cor:genmps}
  The following process generates precisely the $F' \in \Mps$:
  \begin{enumerate}
  \item Choose $F \in \Musat$.
  \item Choose a clause $P$ with $\var(P) \cap \var(F) = \es$ (``P'' like ``pure'').
  \item Choose a map $e: F \ra \pot(P)$ (``e'' like ``extension'').
  \item Let $F' := \set{C \cup e(C) : C \in F}$.
  \end{enumerate}
\end{corol}

For unsatisfiable clause-sets the set of minimally unsatisfiable sub-clause-sets has been studied extensively in the literature; see \cite{MarquesSilva2012MUS} for a recent overview. The set of subsets which are mps's strengthen this notion (now for all clause-sets):
\begin{defi}\label{def:mps}
  For a clause-set $F \in \Cls$ by $\bmm{\mps(F)} \subset \Cls$ the set of all minimal premise sub-clause-sets is denoted: $\mps(F) := \pot(F) \cap \Mps$.
\end{defi}
We have $\abs{\mps(F)} \le 2^{c(F)}-1$.\footnote{There is a typo in Corollary 4.6 of \cite{Kullmann2007ClausalFormZII}, misplacing the ``$-1$'' into the exponent.} The minimal elements of $\mps(F)$ are $\set{C} \in \mps(F)$ for $C \in F$. Since every prime implicate of a clause-set has some minimal premise sub-clause-set, we get that running through all sub-mps's in a clause-set $F$ and extracting the clauses with the pure literals we obtain at least all prime implicates:
\begin{lem}\label{lem:purecprimec}
  For $F \in \Cls$ the map $F' \in \mps(F) \mapsto \purec(F') \sse \set{C \in \Cl : F \models C}$ covers $\primec_0(F)$ (i.e., its range contains the prime implicates of $F$).
\end{lem}

\begin{examp}\label{exp:moremps}
  Examples where we have more minimal premise sub-clause-sets than prime implicates are given by $F \in \Musat$, where $\primec_0(F) = \set{\bot}$, while in the most extreme case every non-empty subset of $F$ can be a minimal premise sub-clause-set (see Theorem \ref{thm:dopedsmumax}).
\end{examp}

\subsection{Doping clause-sets}
\label{sec:Dopingcls}

``Doping'' is the process of adding a unique new variable to every clause of a clause-set. It enables us to follow the usage of this clause in derivations:
\begin{defi}\label{def:doping}
  For every clause-set $F \in \Cls$ we assume an injection $u^F: F \ra \Va \sm \var(F)$ in the following, assigning to every clause $C$ a different variable $u^F_C$. For a clause $C \in \Cl$ and a clause-set $F \in \Cls$ we then define the \textbf{doping} $\bmm{\doping_F(C)} := C \cup \set{u^F_C} \in \Cl$, while $\bmm{\doping(F)} := \set{\doping_F(C) : C \in F} \in \Cls$.
\end{defi}
Remarks:
\begin{enumerate}
\item In the following we drop the upper index in ``$u^F_C$'', i.e., we just use ``$u_C$''.
\item We have $\doping: \Cls \ra \Sat$.
\item For $F \in \Cls$ we have $n(\doping(F)) = n(F) + c(F)$ and $c(\doping(F)) = c(F)$.
\item For $F \in \Cls$ we have $\purec(\doping(F)) = \purec(F) \cup \set{u_C : C \in F}$.
\end{enumerate}
We are interested in the prime implicates of doped clause-sets. It is easy to see that all doped clauses are themselves essential prime implicates:
\begin{lem}\label{lem:necprcdp}
  For $F \in \Cls$ we have $\doping(F) \sse \primec_0(\doping(F))$, and furthermore all elements of $\doping(F)$ are essential prime implicates.
\end{lem}
\begin{prf}
  Every resolvent of clauses from $\doping(F)$ contains at least two doping variables, and thus the clauses of $\doping(F)$ themselves (which contain only one doping variable) are prime and necessary. \Qed
\end{prf}
Thus by Lemma \ref{lem:necprcls} among all the clause-sets equivalent to $\doping(F)$ this clause-set itself is the smallest. Directly by Lemma \ref{lem:characmps} we get that a clause-set is an mps iff its doped form is an mps:
\begin{lem}\label{lem:mpsdoping}
  For $F \in \Cls$ holds $F \in \Mps \Lra \doping(F) \in \Mps$. Thus the map $F' \in \mps(F) \mapsto \doping(F')$ is a bijection from $\mps(F)$ to $\mps(\doping(F))$.
\end{lem}

For doped clause-sets the surjection of Lemma \ref{lem:purecprimec} is bijective:
\begin{lem}\label{lem:purecprimecdoped}
  Consider a clause-set $F \in \Cls$, and let $G := \doping(F)$.
  \begin{enumerate}
  \item\label{lem:purecprimecdoped1} The map $F' \in \mps(G) \mapsto \purec(F') \in \Cl$ is a bijection from $\mps(G)$ to $\primec_0(G)$.
  \item\label{lem:purecprimecdoped2} The inverse map from $\primec_0(G)$ to $\mps(G)$ obtains from $C \in \primec_0(G)$ the clause-set $F' \in \mps(G)$ with $\purec(F') = C$ as $F' = \set{\doping(D) : D \in F \und u_D \in \var(C)}$.
  \end{enumerate}
\end{lem}
\begin{prf}
  By  Lemma \ref{lem:purecprimec} it remains to show that the map of Part \ref{lem:purecprimecdoped1} is injective and does not have subsumptions in the image. Assume for the sake of contradiction there are $G', G'' \in \mps(G)$, $G' \not= G''$, with $\purec(G') \sse \purec(G'')$. Since every clause of $F$ has a different doping-variable, $G' \subset G''$ must hold. Consider the $F', F'' \in \mps(F)$ with $\doping(F')= G'$ and $\doping(F'') = G''$. We have $F' \subset F''$, and thus $\purec(F') \not\sse \purec(F'')$, since for every $F \in \Mps$ the clause $\purec(F)$ is a prime implicate of $F$. It follows that $\purec(G') \not\sse \purec(G'')$, contradicting the assumption. \Qed
\end{prf}

By Lemma \ref{lem:mpsdoping} and Lemma \ref{lem:purecprimecdoped} we obtain:
\begin{thm}\label{thm:dopedmpsprc}
  Consider $F \in \Cls$. Then the map $F' \in \mps(F) \mapsto \purec(\doping(F')) \in \Cl$ is a bijection from $\mps(F)$ to $\primec_0(\doping(F))$.
\end{thm}
Theorem \ref{thm:dopedmpsprc} together with the description of the inversion map in Lemma \ref{lem:purecprimecdoped} yields computation of the set $\mps(F)$ for $F \in \Cls$ via computation of $\primec_0(\doping(F))$.

\begin{corol}\label{cor:surjprimeFG}
  For $F \in \Cls$ we obtain a map from $\primec_0(\doping(F))$ to the set of implicates of $F$ covering $\primec_0(F)$ by the mapping $C \in \primec_0(\doping(F)) \mapsto C \sm V$ for $V := \set{u_C : C \in F}$.
\end{corol}
\begin{prf}
  The given map can be obtained as a composition as follows: For $C \in \primec_0(\doping(F))$ take (the unique) $F' \in \mps(F)$ with $\purec(\doping(F')) = C$, and we have $C \sm V = \purec(F')$. \Qed
\end{prf}

\subsection{Hardness of doped clause-sets}
\label{sec:hddopedcls}

The hardness of a doped clause-set is the maximal hardness of sub-clause-sets of the original clause-set:
\begin{lem}\label{lem:hddoping}
  For $F \in \Cls$ we have $\hardness(\doping(F)) = \max_{F' \sse F} \hardness(F')$.
\end{lem}
\begin{prf}
  We have $\hardness(F') \le \hardness(\doping(F))$ for all $F' \sse F$, since via applying a suitable partial assignment we obtain $F'$ from $F$, setting the doping-variables in $F'$ to false, and the rest to true. And if we consider an arbitrary partial assignment $\vp$ with $\vp * \doping(F) \in \Usat$, then w.l.o.g.\ all doping variables are set (we can set the doping-variables not used by $\vp$ to true, since these variables are all pure), and then we have a partial assignment making $F'$ unsatisfiable for that $F' \in \Usat$ given by all the doping variables set by $\vp$ to false. \Qed
\end{prf}

\begin{examp}\label{exp:hddot}
  For an example of a clause-set $F \in \Usat$ with $\hardness(\doping(F)) > \hardness(F)$ consider any clause-set $F' \in \Cls$ with $\hardness(F') > 0$, and then take $F := F' \cup \set{\bot}$ (note that $\bot \notin F'$). Thus $\hardness(F) = 0$. And by Part 1 of Lemma 6.5 in \cite{GwynneKullmann2012Slur,GwynneKullmann2012SlurJ}, all $\Urefc_k$ are closed under partial assignments, so for $\vp := \pao{u_\bot}{1} \cup \pab{u_C \ra 0 \mb C \in F'}$ we have $\hardness(\doping(F)) \ge \hardness(\vp * \doping(F)) = \hardness(F') > \hardness(F) = 0$.
\end{examp}

\section{Doping tree clause-sets}
\label{sec:doptreecls}

As explained in Subsection \ref{sec:introstricteq}, we want to construct boolean functions (given by clause-sets) with a large number of prime implicates, where we have strong control over these prime implicates. For this purpose we dope ``minimally unsatisfiable clause-sets of deficiency $1$'', that is the elements of $\Smusati{\delta=1}$. First we review in Subsection \ref{sec:prelimMU} the background (for more information see \cite{Kullmann2007HandbuchMU}). In Subsection \ref{sec:totalmps} we show that these clause-sets are the core of ``total minimal premise sets'', which have as many minimal-premise sub-clause-sets as possible. In Theorem \ref{thm:dopedsmumax} we show that $F \in \Smusati{\delta=1}$ are precisely the unsatisfiable clause-sets such that every non-empty subset is an mps. Then in Subsection \ref{sec:appSMU1dop} we consider doping of these special clause-sets, and in Theorem \ref{thm:sumdsmuo} we determine basic properties of $\doping(F)$.

\subsection{Preliminaries on minimal unsatisfiability}
\label{sec:prelimMU}

A minimally unsatisfiable $F \in \Musat$ is \emph{saturated minimally unsatisfiable} iff for all clauses $C \in F$ and for every literal $x$ with $\var(x) \notin \var(C)$ the clause-set $(F \sm C) \cup (C \cup \set{x})$ is satisfiable. The set of all saturated minimally unsatisfiable clause-sets is denoted by $\bmm{\Smusat} \subset \Musat$. By \bmm{\Smusati{\delta=k}} we denote the set of $F \in \Smusat$ with $\delta(F) = k$, where the \emph{deficiency} of a clause-set $F$ is given by $\delta(F) := c(F) - n(F)$. In \cite{Ku99dKo} (generalised in \cite{Kullmann2007ClausalFormZII}) it is shown that the elements of $\Smusati{\delta=1}$ are exactly the clause-sets introduced in \cite{Co73}. The details are as follows. For rooted trees $T$ we use \bmm{\nds(T)} for the set of nodes and $\bmm{\lvs(T)} \sse \lvs(T)$ for the set of leaves, and we set $\bmm{\nnds(T)} := \abs{\nds(T)}$ and $\bmm{\nlvs(T)} := \abs{\lvs(T)}$. In our context, the nodes of rooted trees are just determined by their positions, and do not have names themselves. Another useful notation for a tree $T$ and a node $w$ is \bmm{T_w}, which is the sub-tree of $T$ with root $w$; so $\lvs(T) = \set{w \in \nds(T) : \nnds(T_w) = 1}$. Recall that for a full binary tree $T$ (every non-leaf node has two children) we have $\nnds(T) = 2 \nlvs(T) - 1$.

\begin{defi}\label{def:vardisjointreerep}
  Consider a full binary tree $T$ and an injective vertex labelling $u : (\nds(T) \sm \lvs(T)) \ra \Va$ for the inner nodes; the set of all such pairs is denoted by \bmm{\Tsmuo}. The induced edge-labelling assigns to every edge from an inner node $w$ to a child $w'$ the literal $u(w)$ resp.\ $\ol{u(w)}$ for a left resp.\ right child. We define the \textbf{clause-set representation $\bmm{\smuo(T,u)}$} (where ``1'' reminds of deficiency $1$ here; see Lemma \ref{lem:charsmu}) to be $\bmm{\smuo(T,u)} := \set{ C_w : w \in \lvs(T) }$, where clause \bmm{C_w} consists of all the literals (i.e., edge-labels) on the path from the root of $T$ to $w$.
\end{defi}
By Lemma C.5 in \cite{Ku99dKo} we know that via this tree-construction we obtain exactly the clause-sets in $\Smusati{\delta=1}$:
\begin{lem}\label{lem:charsmu} $\smuo: \Tsmuo \ra \Smusati{\delta=1}$ is a bijection.
\end{lem}
By $\bmm{\tsmuo}: \Smusati{\delta=1} \ra \Tsmuo$ we denote the inversion of $\smuo$. Typically we identify $(T,u) \in \Tsmuo$ with $T$, and let the context determine $u$. So $\tsmuo(F)$ is the full binary tree, where the variable $v$ labelling the root (for $F \not= \set{\bot}$) is the unique variable occurring in every clause of $F$, and the clause-sets determining the left resp.\ right subtree are $\pao v0 * F$ resp.\ $\pao v1 * F$. By $\bmm{w_C}$ for $C \in F$ we denote the leaf $w$ of $\tsmuo(F)$ such that $C_w = C$. Furthermore we identify the literals of $F$ with the edges of $\tsmuo(F)$. Note that $c(F) = \nlvs(\tsmuo(F))$ and $n(F) = \nnds(\tsmuo(F)) - \nlvs(\tsmuo(F))$.

\begin{examp}\label{exp:dopedsmutree}
  Consider the following labelled binary tree $T$ (using additionally labels $1,\dots,6$ for the leaves):
  \begin{displaymath}
    \xygraph{
      !{0;/r8ex/:}
        []{v_1} (
          - [dll]{v_2}_{v_1} (
            -[dll]{v_3}_{v_2} (
              -[dl]{1}_{v_3} (),
              -[dr]{2}^{\ol{v_3}} ()
            ),
            -[drr]{v_4}^{\ol{v_2}} (
              -[dl]{3}_{v_4} (),
              -[dr]{4}^{\ol{v_4}} ()
            )
          ),
          - [drr]{v_5}^{\ol{v_1}} (
            -[dl]{5}_{v_5} (),
            -[dr]{6}^{\ol{v_5}} ()
          )
        )
    }
  \end{displaymath}
  Then $\smuo(T) = \set{ \set{v_1,v_2,v_3}, \set{v_1,v_2,\ol{v_3}}, \set{v_1,\ol{v_2},v_4}, \set{v_1,\ol{v_2},\ol{v_4}}, \set{\ol{v_1},v_5}, \set{v_1,\ol{v_5}} }$, where for example $C_3 = \set{v_1,\ol{v_2},v_4}$ and $w_{\set{v_1,\ol{v_5}}} = 6$.
\end{examp}

We note in passing, that those $\smuo(T)$ with $\hts(T) \le 1$ can be easily characterised as follows. A clause $C \in F$ for $F \in \Cls$ is called \emph{full} if $\var(C) = \var(F)$, that is, $C$ contains all variables of $F$.
\begin{lem}\label{lem:characsmuhts1}
  $F \in \Smusati{\delta=1}$ contains a full clause if and only if $\hts(\tsmuo(F)) \le 1$.
\end{lem}
See Example \ref{exp:sepUC01} for more on these special clause-sets. The effect of applying a partial assignment to some element of $\Smusati{\delta=1}$ is easily described as follows:
\begin{lem}\label{lem:apppasmuo}
  Consider $F \in \Smusati{\delta=1}$ and $x \in \lit(F)$, and let $F' := \pao x1 * F$. We have:
  \begin{enumerate}
  \item $F' \in \Smusati{\delta=1}$.
  \item Let $T := \tsmuo(F)$ and $T' := \tsmuo(F')$. The tree $T'$ is obtained from $T$ as follows:
    \begin{enumerate}
    \item Consider the node $w \in T$ labelled with $\var(x)$. Let $T_x, T_{\ol{x}}$ be the two subtrees hanging at $w$, following the edge labelled with $x$ resp.\ $\ol{x}$.
    \item Now $T'$ is obtained from $T'$ by removing subtree $T_x$, and attaching $T_{\ol{x}}$ directly at position $w$.
    \end{enumerate}
  \end{enumerate}
\end{lem}

\begin{examp}\label{exp:apppasmuo}
  Consider the labelled binary tree $T$ from Example \ref{exp:dopedsmutree} where
  \begin{displaymath}
    \smuo(T) = \set{ \underbrace{\set{v_1,v_2,v_3}}_{\bmm{C_1}}, \underbrace{\set{v_1,v_2,\ol{v_3}}}_{\bmm{C_2}}, \underbrace{\set{v_1,\ol{v_2},v_4}}_{\bmm{C_3}}, \underbrace{\set{v_1,\ol{v_2},\ol{v_4}}}_{\bmm{C_4}}, \underbrace{\set{\ol{v_1},v_5}}_{\bmm{C_5}}, \underbrace{\set{v_1,\ol{v_5}}}_{\bmm{C_6}} }
  \end{displaymath}
  Now consider the application of the partial assignment $\pao {v_2}1$ to $\smuo(T)$:
  \begin{enumerate}
  \item Clauses $C_1$ and $C_2$ are satisfied, and so are removed (both contain $v_2$).
  \item Clauses $C_3$ and $C_4$ both contain $\ol{v_2}$ and so this literal is removed.
  \end{enumerate}
  This yields:
  \begin{displaymath}
    \pao {v_2}1 * \smuo(T) = \set{ \underbrace{\set{v_1,v_4}}_{\bmm{C_3 \sm \set{\ol{v_2}}}}, \underbrace{\set{v_1,\ol{v_4}}}_{\bmm{C_4 \sm \set{\ol{v_2}}}}, \underbrace{\set{\ol{v_1},v_5}}_{\bmm{C_5}}, \underbrace{\set{v_1,\ol{v_5}}}_{\bmm{C_6}} }
  \end{displaymath}
  The satisfaction (removal) of clauses and removal of literals is illustrated directly on $T$ in Figure \ref{fig:dopedsmutreepass} with dotted and dashed lines for clause and literal removal respectively. The tree corresponding to $\pao {v_2}1 * \smuo(T)$ is illustrated in Figure \ref{fig:dopedsmutreepassafter}.
  \begin{figure}[ht]
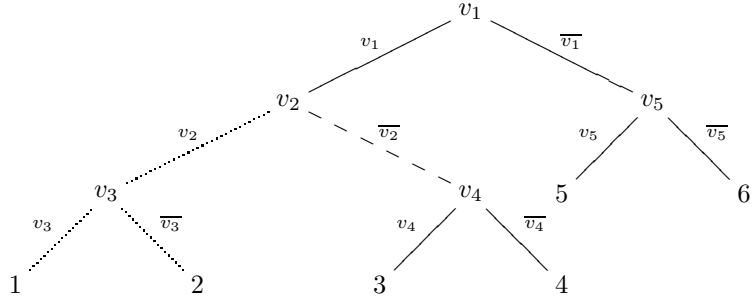

    \begin{displaymath}
    \xygraph{
      !{0;/r8ex/:}
        []{v_1} (
          - [dll]{v_2}_{v_1} (
            -@{..}[dll]{v_3}_{v_2} (
              -@{..}[dl]{1}_{v_3} (),
              -@{..}[dr]{2}^{\ol{v_3}} ()
            ),
            -@{--}[drr]{v_4}^{\ol{v_2}} (
              -[dl]{3}_{v_4} (),
              -[dr]{4}^{\ol{v_4}} ()
            )
          ),
          - [drr]{v_5}^{\ol{v_1}} (
            -[dl]{5}_{v_5} (),
            -[dr]{6}^{\ol{v_5}} ()
          )
        )
    }
  \end{displaymath}
  \caption{Illustration of application of $\pao {v_2}1$ to $\smuo(T)$. Dotted lines indicate that the clauses corresponding to the effected leaves are satisfied; dashed lines indicate that the corresponding literal is falsified and therefore removed from all clauses.}
  \label{fig:dopedsmutreepass}
  \end{figure}
  \begin{figure}
    \begin{displaymath}
    \xygraph{
      !{0;/r8ex/:}
        []{v_1} (
          - [dll]{v_4}_{v_1} (
            -[dl]{3}_{v_4} (),
            -[dr]{4}^{\ol{v_4}} ()
          ),
          - [drr]{v_5}^{\ol{v_1}} (
            -[dl]{5}_{v_5} (),
            -[dr]{6}^{\ol{v_5}} ()
          )
        )
    }
  \end{displaymath}
  \caption{Tree associated with $\pao {v_2}1 * \smuo(T)$.}
  \label{fig:dopedsmutreepassafter}
  \end{figure}
\end{examp}

\begin{corol}\label{cor:SMU1stable}
  $\Smusati{\delta=1}$ is stable under application of partial assignments, that is, for $F \in \Smusati{\delta=1}$ and $\vp \in \Pass$ holds $\vp * F \in \Smusati{\delta=1}$.
\end{corol}

From Lemma \ref{lem:charsmu} follows $\Smusati{\delta=1} \subset \Uclash$, where $\bmm{\Clash} \subset \Cls$ is the set of \emph{hitting clause-sets}, that is, those $F \in \Cls$ where every two clauses clash in at least one literal, i.e., for all $C, D \in F$, $C \not= D$, we have $\abs{C \cap \ol{D}} \ge 1$, and $\bmm{\Uclash} := \Clash \cap \Usat$. It is well-known that $\Uclash \subset \Smusat$ holds (for a proof see Lemma 2 in \cite{KullmannZhao2012ConfluenceJ}).

\subsection{Total minimal premise sets}
\label{sec:totalmps}

We are interested in clause-sets which have as many sub-mps's as possible:
\begin{defi}\label{def:tmps}
  A clause-set $F \not= \top$ is a \textbf{total mps} if $\mps(F) = \pot(F) \sm \set{\top}$.
\end{defi}
Every total mps is an mps.
\begin{examp}\label{exp:totalmps}
  $\set{\set{a,b},\set{\ol{a},b},\set{\ol{b}}}$ is a total mps, while $\set{\set{a,b},\set{\ol{a}},\set{\ol{b}}}$ is an mps (since minimally unsatisfiable), but not a total mps.
\end{examp}

To determine all total mps's, the central task to determine the minimally unsatisfiable total mps's. Before we can prove that these are precisely the saturated minimally unsatisfiable clause-sets of deficiency $1$, we need to state a basic property of these clause-sets, which follows by definition of $\tsmuo(F)$ for $F \in \Smusati{\delta=1}$ (recall Subsection \ref{sec:prelimMU}):
\begin{lem}\label{lem:detpureSMU1}
  Consider $F \in \Smusati{\delta=1}$ and $F' \sse F$. Let $T := \tsmuo(F)$. The set $\purec(F')$ of pure literals of $F'$ can be determined as follows:
  \begin{enumerate}
  \item Let $W_{F'} := \set{w_C : C \in F'} \sse \lvs(T)$ be the set of leaves corresponding to the clauses of $F'$.
  \item For a literal $x \in \lit(F)$ let $w \in \nds(T)$ be the node labelled with $\var(x)$, and let $T_x$ the the subtree of $w$ reached by $x$, and let $T_{\ol{x}}$ be the subtree of $w$ reached by $\ol{x}$.
  \item Now $x \in \purec(F')$ if and only if $W_{F'} \cap \lvs(T_x) \not= \es$ and $W_{F'} \cap \lvs(T_{\ol{x}}) = \es$.
  \end{enumerate}
\end{lem}

\begin{examp}\label{exp:purecSMU1}
  Consider the clause-set
  \begin{multline*}
    F := \setb{ \underbrace{\set{v_1,v_2,v_3}}_{\bmm{C_1}}, \underbrace{\set{v_1,v_2,\ol{v_3}}}_{\DST C_2}, \underbrace{\set{v_1,\ol{v_2}, v_4}}_{\bmm{C_3}}, \underbrace{\set{v_1, \ol{v_2}, \ol{v_4}}}_{\bmm{C_4}},\\
      \underbrace{\set{\ol{v_1}, v_5, v_6}}_{\DST C_5}, \underbrace{\set{\ol{v_1}, v_5, \ol{v_6}}}_{\DST C_6}, \underbrace{\set{\ol{v_1}, \ol{v_5}}}_{\bmm{C_7}} }
  \end{multline*}
  and the subset $F' := \set{C_1, C_3, C_4, C_7}$. The tree $\tsmuo(F)$ is as follows, with the dashed edges representing literals not in $\bc F' = \set{v_1,v_2,v_3,v_4,\ol{v_1},\ol{v_2},\ol{v_4},\ol{v_5}}$:
  \begin{displaymath}
    \xygraph{
      !{0;/r7ex/:}
        []{v_1} (
          - [dlll]{v_2}_{v_1} (
            -[dll]{v_3}_{v_2} (
              -[dl]{\bmm{1}}_{v_3} (),
              -@{--}[dr]{2}^{\ol{v_3}} ()
            ),
            -[drr]{v_4}^{\ol{v_2}} (
              -[dl]{\bmm{3}}_{v_4} (),
              -[dr]{\bmm{4}}^{\ol{v_4}} ()
            )
          ),
          - [drrr]{v_5}^{\ol{v_1}} (
            -@{--}[dl]{v_6}_{v_5} (
              -@{--}[dl]{5}_{v_6} (),
              -@{--}[dr]{6}^{\ol{v_6}} ()
            ),
            -[dr]{\bmm{7}}^{\ol{v_5}} ()
          )
        )
    }
  \end{displaymath}
  We have $W_{F'} = \set{1,3,4,7}$ and
  \begin{displaymath}
    \purec(F') = \bc F' \sm \set{\underbrace{v_2,\ol{v_2}}_{\bmm{C_1, C_3}}, \underbrace{v_1,\ol{v_1}}_{\bmm{C_1, C_7}},\underbrace{v_4,\ol{v_4}}_{\bmm{C_3, C_4}}} = \set{v_3, \ol{v_5}}.
  \end{displaymath}
  Now consider $x \in \lit(F)$:
  \begin{enumerate}
  \item For $x = v_3$ holds $\lvs(T_{v_3}) \cap W_{F'} = \set{1}$ and $T_{\ol{v_3}} \cap W_{F'} = \es$, thus $v_3 \in \purec(F')$.
  \item For $x = \ol{v_5}$ holds $\lvs(T_{\ol{v_5}}) \cap W_{F'} = \set{7}$ and $T_{v_5} \cap W_{F'} = \es$, thus $\ol{v_5} \in \purec(F')$.
  \item Considering for example $x = v_1$, we have $\lvs(T_{v_1}) \cap W_{F'} = \set{1,3}$ and $\lvs(T_{\ol{v_1}}) \cap W_{F'} = \set{7}$, thus $v_1 \notin \purec(F')$, while for $x = v_6$ we have $\lvs(T_{v_6}) \cap W_{F'} = \es$ and $\lvs(T_{\ol{v_6}}) \cap W_{F'} = \es$, thus $v_6 \notin \purec(F')$.
  \end{enumerate}
\end{examp}

\begin{thm}\label{thm:dopedsmumax}
  An unsatisfiable clause-set $F \in \Usat$ is a total mps if and only if $F \in \Smusati{\delta=1}$.
\end{thm}
\begin{prf}
  First assume that $F$ is a total mps. Then every two clauses $C, D \in F$, $C \not= D$, clash in exactly one literal (otherwise $\set{C,D} \notin \Mps$). In \cite{Ku2003e}, Corollary 34, it was shown that that an unsatisfiable clause-sets $F$ has precisely one clash between any pair of different clause-sets iff $F \in \Smusati{\delta=1}$ holds (an alternative proof was found in \cite{SloanSzoerenyiTuran2005Primimplikanten_1}).\footnote{In \cite{Ku2003e} the notation ``$\Uclash$'' was used to denote ``uniform hitting clause-sets'', which is now more appropriately called ``(conflict-)regular hitting clause-sets'', while ``U'' now stands for ``unsatisfiable''.} Now assume $F \in \Smusati{\delta=1}$, and we have to show that $F$ is a total mps. So consider $F' \in \pot(F) \sm \set{\top}$, and let $C := \purec(F)$, $\vp := \vp_C$. Since $F'$ is a hitting clause-set, $\vp$ is contraction-free for $F'$, and according to Lemma \ref{lem:characmps} it remains to show that $F'' := \vp * F'$ is unsatisfiable (recall that hitting clause-sets are irredundant). Assume that $F''$ is satisfiable, and consider a partial assignment $\psi$ with $\psi * F'' = \top$ and $\var(\psi) \cap \var(\vp) = \es$. We show that then $\vp \cup \psi$ would be a satisfying assignment for $F$, contradicting the assumption. To this end it suffices to show that for all $D \in F \sm F'$ holds $\ol{C} \cap D \not= \es$. Consider $T := \tsmuo(F)$, and let $W_{F'}$ be defined as in Lemma \ref{lem:detpureSMU1}. Starting from the leaf $w_D$, let $w$ be the first node on the path to the root of $T$ such that one of the two subtrees of $w$ contains a leaf of $W_{F'}$. Let $\ol{x}$ be the literal at $w$ on the path to $w_D$. So by Lemma \ref{lem:detpureSMU1} we have $x \in C$, while by definition $\ol{x} \in D$. \Qed
\end{prf}

\begin{corol}\label{cor:alltotmps}
  For a clause-set $F \in \Cls$ the following properties are equivalent:
  \begin{enumerate}
  \item $F$ is a total mps.
  \item $\vp_{\purec(F)} * F \in \Smusati{\delta=1}$, and $\vp_{\purec(F)}$ is contraction-free for $F$.
  \end{enumerate}
\end{corol}
\begin{prf}
Let $F' := \vp_{\purec(F)} * F$. If $F$ is a total mps, then by Lemma \ref{lem:characmps} follows $F' \in \Musat$, where $\vp_{\purec(F)}$ is contraction-free for $F$. Also by Lemma \ref{lem:characmps} follows then, that $F' \in \Mps$, and thus by Theorem \ref{thm:dopedsmumax} we obtain $F' \in \Smusati{\delta=1}$. For the other direction, if $F' \in \Smusati{\delta=1}$ holds, where $\vp_{\purec(F)}$ is contraction-free for $F$, then by Theorem \ref{thm:dopedsmumax} follows that $F'$ is a total mps, which by Lemma \ref{lem:characmps} yields that $F$ is a total mps.  \Qed
\end{prf}

Thus we can precisely construct all total mps's, if we start the process described in Corollary \ref{cor:genmps} not with an arbitrary $F \in \Musat$, but with an $F \in \Smusati{\delta=1}$.

\begin{examp}\label{exp:countertmps}
  That every $2$-element sub-clause-set of $F \in \Cls$ is an mps, that is, every two (different) clauses of $F$ clash in precisely one literal, says that $F$ is $1$-regular hitting in the terminology of \cite{Kullmann2007ClausalFormZII}, Section 6. For $F \in \Usat$ the proof of Theorem \ref{thm:dopedsmumax} shows, that $F$ is a total mps iff $F$ is $1$-regular hitting. However for $F \in \Sat$ this is not true, and the simplest example is $F := \set{\set{\ol{a},b},\set{\ol{b},c},\set{\ol{c},a}}$: $F$ is $1$-regular hitting, but has no pure literal and is satisfiable, and thus $F \notin \Mps$. In this case we have $\delta(F) = 0$. For an interesting example with deficiency $1$ see Section 5 in \cite{Ku2003e}.
\end{examp}
We arrive at a simple and perspicuous proof of the main result of \cite{SloanSzoerenyiTuran2005Primimplikanten_1}, that the clause-sets $F$ with $\abs{\primec_0(F)} = 2^{c(F)} - 1$ are precisely the clause-sets $\doping(F)$ for $F \in \Smusati{\delta=1}$ when allowing to replace the single doping variable of a clause by any non-empty set of new (pure) literals:
\begin{lem}\label{lem:characmaxnprimeimpl}
  For $F \in \Cls \sm \set{\top}$ holds $\abs{\primec_0(F)} = 2^{c(F)} - 1$ if and only if the following two conditions hold:
  \begin{enumerate}
  \item $F$ is a total mps.
  \item For every clause $C \in F$ there is $x \in C$ such that $\var(x) \notin \var(F \sm \set{C})$.
  \end{enumerate}
\end{lem}
\begin{prf}
First assume $\abs{\primec_0(F)} = 2^{c(F)} - 1$. Thus the map $F' \in \mps(F) \mapsto \purec(F') \sse \set{C \in \Cl : F \models C}$, which according to Lemma \ref{lem:purecprimec} covers $\primec_0(F)$, must indeed be a bijection from $\mps(F)$ to $\primec_0(F)$, and hence $F$ is a total mps (here we need $F \ne \top$). If there would be $C \in F$ such that for all $x \in C$ we have $\var(x) \in \var(F \sm \set{C})$, then $\purec(F) \sse \purec(F \sm \set{C})$, and thus $F \sm \set{C}$ could not yield a prime implicate different from the prime implicate obtained from $F$.

The inverse direction follows by the observation, that the existence of the unique ``doping literals'' $x \in C$ has the consequence, that for $\top \subset F', F'' \sse F$ with $F' \ne F''$ we get $\purec(F') \ne \purec(F'')$, since these doping literals make a difference. \Qed
\end{prf}

\subsection{Doping $\Smusati{\delta=1}$}
\label{sec:appSMU1dop}

We are turning now our attention to a closer understanding of the prime implicates $C$ of doped $F \in \Smusati{\delta=1}$. We start with their identification with non-empty sub-clause-sets $F'$ of $\doping(F)$:
\begin{lem}\label{lem:dopedsmumax2}
  Consider a clause-set $F \in \Smusati{\delta=1}$. By Theorem \ref{thm:dopedsmumax} each non-empty subset yields a minimal premise set. Thus by Theorem \ref{thm:dopedmpsprc} we have:
  \begin{enumerate}
  \item $\primec_0(\doping(F)) = \set{ \purec(F') \mb \top \not= F' \sse \doping(F) }$.
  \item $\abs{\primec_0(\doping(F))} = 2^{c(F)} - 1$.
  \end{enumerate}
\end{lem}
Since the clauses of $\doping(F)$ can be identified with leaves of the tree $\tsmuo(F)$, we obtain a bijection between non-empty sets $V$ of leaves of the tree $\tsmuo(F)$ and prime implicates of $\doping(F)$:
\begin{defi}\label{def:clausesDSMU1}
  For $F \in \Smusati{\delta = 1}$ and $\es \not= V \sse \lvs(\tsmuo(F))$ the clause \bmm{C_V} is the prime implicate $\purec(\set{C_w \in F \mb w \in V})$ of $\doping(F)$ according to Lemma \ref{lem:dopedsmumax2}. For $w \in \lvs(\tsmuo(F))$ we furthermore set $\bmm{u_w} := u_{C_w}$.
\end{defi}

By Lemma \ref{lem:dopedsmumax2}:
\begin{lem}\label{lem:dopedsmuprimetreechar}
  For $F \in \Smusati{\delta = 1}$ holds $\primec_0(\doping(F)) = \set{ C_V \mb \es \not= V \sse \lvs(\tsmuo(F)) }$.
\end{lem}

How precisely from $V \sse \lvs(\tsmuo(F))$ the prime implicate $C_V$ is constructed shows the following lemma:
\begin{lem}\label{lem:dopedsmuprimechar}
  Consider $F \in \Smusat_{\delta=1}$ and $\es \not= V \sse \lvs(\tsmuo(F))$. We have $C_V = U_V \cup P_V$, $U_V \cap P_V = \es$, where
  \begin{enumerate}
  \item $U_V := \set{u_w \mb w \in V}$, and
  \item $P_V := \purec(F')$ for $F' := \set{C_w : w \in V}$ as given in Lemma \ref{lem:detpureSMU1}, that is, $P_V$ is the set of literals $x$ such that $V \cap \lvs(T_x) \not= \es$ and $V \cap \lvs(T_{\ol{x}}) = \es$.
  \end{enumerate}
\end{lem}

\begin{examp}\label{exp:CV}
  Consider the clause-set
  \begin{displaymath}
    F := \set{ \set{v_1,v_2}, \set{v_1,\ol{v_2}}, \set{\ol{v_1},v_3}, \set{\ol{v_1},\ol{v_3}} } \in \Smusat_{\delta=1}
  \end{displaymath}
  corresponding to the tree
  \begin{displaymath}
    \xygraph{
      !{0;/r8ex/:}
        []{v_1} (
          - [dll]{v_2}_{v_1} (
            -[dl]{\bmm{1}}_{v_2} (),
            -[dr]{2}^{\ol{v_2}} ()
          ),
          - [drr]{v_3}^{\ol{v_1}} (
            -[dl]{\bmm{3}}_{v_3} (),
            -[dr]{4}^{\ol{v_3}} ()
          )
        )
    }
  \end{displaymath}
  with the doped clause-set
  \begin{displaymath}
    \doping(F) = \set{ \set{v_1,v_2,u_1}, \set{v_1,\ol{v_2},u_2}, \set{\ol{v_1},v_3,u_3}, \set{\ol{v_1},\ol{v_3}, u_4 }}.
  \end{displaymath}
  Now consider the set $V := \set{1,3}$. According to Definition \ref{def:clausesDSMU1} we have that $C_V = \purec(\set{ \set{v_1,v_2,u_1}, \set{\ol{v_1},v_3,u_3} }) = \set{v_2,v_3,u_1,u_3}$. By Lemma \ref{lem:dopedsmuprimechar} we have that $C_V = U_V \cup P_V$, where $U_V = \set{u_1, u_3}$ and $P_V = \purec( \set{ \set{v_1,v_2}, \set{\ol{v_1},v_3} } = \set{v_2,v_3}$. Note that for both $x \in \set{v_2,v_3} = P_V$ we have that $\lvs(T_{x}) \cap V \not= \es$ and $\lvs(T_{\ol{x}}) \cap V = \es$, but we do not have this for $x \in \lit(F) \sm \set{v_2,v_3}$.
\end{examp}

The hardness of $F$ as well as $\doping(F)$ is the Horton-Strahler number of $\tsmuo(F)$:
\begin{lem}\label{lem:hdsmu1}
  Consider $F \in \Smusati{\delta=1}$, and let $k := \hts(\tsmuo(F))$. Then we have $\hardness(F) = \hardness(\doping(F)) = k$.
\end{lem}
\begin{prf}
Let $T := \tsmuo(F)$. First we show $\hardness(F) = k$. We have $\hardness(F) \le k$, since $T$ is by definition of $F = \smuo(T)$ already a resolution tree (when extending the labelling of leaves to all nodes), deriving $\bot$ from $F$. To show $\hardness(F) \ge k$, we use Lemma \ref{lem:lbhd} with $\mc{C} := \Smusati{\delta=1}$ and $h(F) := \hts(\tsmuo(F))$. Based on Lemma \ref{lem:apppasmuo}, we consider the effect on the Horton-Strahler number of assigning a truth value to one variable $v \in \var(F)$. Let $w \in \nds(T)$ be the (inner) node labelled with $v$, and let $T^w_0, T^w_1$ be the left resp.\ right subtree hanging at $w$. Now the effect of assigning $\ve \in \set{0,1}$ to $v$ is to replace $T_w$ with $T^w_{\ve}$. Let $T_{\ve}$ be the (whole) tree obtained by assigning $\ve$ to $v$, that is, $T_{\ve} := \tsmuo(\pao {v}{\ve} * F)$. If $\hts(T^w_0) = \hts(T^w_1)$, then we have $\hts(T_{\ve}) \ge k-1$, since at most one increase of the Horton-Strahler number for subtrees is missed out now. Otherwise we have $\hts(T_0) = \hts(T)$ or $\hts(T_1) = \hts(T)$, since removal of the subtree with the smaller Horton-Strahler number has no influence on the Horton-Strahler number of the whole tree. So altogether Lemma \ref{lem:lbhd} is applicable, which concludes the proof of $\hardness(F) = k$.

For showing $\hardness(\doping(F)) = k$ we use Lemma \ref{lem:hddoping}: so consider $F' \sse F$ and $\vp \in \Pass$ with $\vp * F' \in \Usat$, let $F'' := \vp * F'$, and we have to show $\hardness(F'') \le k$. W.l.o.g.\ $\var(\vp) \sse \var(F')$. By Corollary \ref{cor:SMU1stable} we have that $\vp *F \in \Smusati{\delta=1}$, and thus $\vp *F = F''$ must hold, and $\hardness(F'') = \hts(\tsmuo(F''))$ (by the first part). By Lemma \ref{lem:apppasmuo}, $\tsmuo(F'')$ results from $T$ by a sequence of removing subtrees, and it is easy to see, that thus $\hts(\tsmuo(F'')) \le k$ holds. \Qed
\end{prf}

We summarise what we have learned about $\doping(F)$ for $F \in \Smusati{\delta=1}$:
\begin{thm}\label{thm:sumdsmuo}
  Consider $F \in \Smusati{\delta=1}$.
  \begin{enumerate}
  \item For each clause-set $F'$ equivalent to $\doping(F)$ there is an injection $i: \doping(F) \ra F'$ with $\fa\, C \in \doping(F) : C \sse i(C)$ (by Lemma \ref{lem:necprcdp}).
  \item $\doping(F)$ is a total mps (by Corollary \ref{cor:alltotmps}).
  \item The prime implicates of $\doping(F)$ are given by Lemmas \ref{lem:dopedsmuprimetreechar}, \ref{lem:dopedsmuprimechar}.
  \item $\hardness(\doping(F)) = \hts(\tsmuo(F))$ (by Lemma \ref{lem:hdsmu1}).
  \end{enumerate}
\end{thm}

\section{Separating $\Urefc_{k+1}$ from $\Wrefc_k$}
\label{sec:lowerb}

This section proves the main result of this article, Theorem \ref{thm:separation}, which exhibits for every $k \ge 0$ sequences $(F^k_h)_{h \in \NN}$ of small clause-sets of hardness $k+1$, where every equivalent clause-set of hardness $k$ (indeed of w-hardness $k$) is of exponential size. In this way we show that the $\Urefc_k$ hierarchy as well as the $\Wrefc_k$ hierarchy is useful, i.e., equivalent clause-sets with higher (w-)hardness can be substantially shorter. These $F^k_h$ are doped versions of clause-sets from $\Smusati{\delta=1}$ (recall Theorem \ref{thm:sumdsmuo}), which are ``extremal'', that is, their underlying trees $\tsmuo(F^k_h)$ are for given Horton-Strahler number $k+1$ and height $h$ as large as possible.

The organisation of this section is as follows: In Subsection \ref{sec:triggerhyp} the main tool for showing size-lower-bounds for equivalent clause-sets of a given (w-)hardness is established in Theorem \ref{thm:triggersetmethod}. Subsection \ref{sec:est} introduces the ``extremal trees''. Subsection \ref{sec:hierbadrep} shows the main lower bound in Theorem \ref{thm:nogoodksoft}, and applies it to show the separation Theorem \ref{thm:separation}.

\subsection{Trigger hypergraphs}
\label{sec:triggerhyp}

Our goal is to construct clause-sets $F^k_h$ of hardness $k+1$, which have no short equivalent clause-set $F$ with $\whardness(F) \le k$, where w.l.o.g.\ $F \sse \primec_0(F^k_h) = \primec_0(F)$. This subsection is about the general lower-bound method. How are we going to find a lower bound on the number of clauses of $F$ ? The property $\whardness(F) \le k$ means, that for every $C \in \primec_0(F)$ the unsatisfiable clause-set $\vp_C * F$ can be refuted by $k$-resolution. In order for $k$-resolution to have a chance, there must be at least one clause of length at most $k$ in $\vp_C * F$ --- and this necessary condition is all we consider. So our strategy is to show that every $F \sse \primec_0(F^k_h)$, such that for all $C \in \primec_0(F^k_h)$ there is a clause of length at most $k$ in $\vp_C * F$, is big.

It is useful to phrase this approach in hypergraph terminology. Recall that a hypergraph is a pair $G = (V,E)$, where $V$ is a set (of ``vertices'') and $E \sse \pot(V)$ (the set of hyperedges), where one uses $V(G) := V$ and $E(G) := E$. A \emph{transversal} of $G$ is a set $T \sse V(G)$ such that for all $E \in E(G)$ holds $T \cap E \not= \es$. The minimum size of a transversal is denoted by \bmm{\tau(G)}, the \textbf{transversal number}.
\begin{defi}\label{def:triggerhypergraph}
  Consider $k \in \NNZ$ and $F \in \Cls$. The \textbf{trigger hypergraph} $\trighyp{F}{k}$ is the hypergraph with the prime implicates of $F$ as its vertices, and for every prime implicate $C$ of $F$ a hyperedge $E^k_C$. The hyperedge $E^k_C$ contains all prime implicates $C' \in \primec_0(F)$ which are not satisfied by $\vp_C$ and yield a clause of size at most $k$ under $\vp_C$. That is,
  \begin{enumerate}
  \item $V(\trighyp{F}{k}) := \primec_0(F)$, and
  \item $E(\trighyp{F}{k}) := \set{E^k_C \mb C \in \primec_0(F)}$,
  \end{enumerate}
  where $E^k_C := \set{C' \in \primec_0(F) \mb C' \cap \ol{C} = \es \und \abs{C' \sm C} \le k}$.
\end{defi}
Note that the trigger hypergraph of $F \in \Cls$ depends only on the underlying boolean function of $F$, and thus for every equivalent $F'$ we have $\trighyp{F'}{k} = \trighyp{F}{k}$.
\begin{examp}\label{exp:trighyp}
  Consider the clause-set
  \begin{displaymath}
    F := \setb{\underbrace{\set{v_1,\ol{v_3},\ol{v_4}}}_{C_1},\underbrace{\set{v_2,v_3,\ol{v_4}}}_{C_2},\underbrace{\set{v_2,\ol{v_3},v_4}}_{C_3},
               \underbrace{\set{\ol{v_2},v_3,v_4}}_{C_4},\underbrace{\set{v_1,v_3,v_4}}_{C_5},\underbrace{\set{v_1,v_2}}_{C_6}}.
  \end{displaymath}
  As shown in Example 8.2 of \cite{GwynneKullmann2012Slur,GwynneKullmann2012SlurJ} we have $\primec_0(F) = F$. The trigger hypergraph $\trighyp{F}{0}$ is (as always) the hypergraph with all singleton sets, i.e., $E(\trighyp{F}{0}) = \setb{\set{C_1},\dots,\set{C_6}}$. The hypergraphs $\trighyp{F}{k}$ for $k \in \set{1,2}$ are represented by Figures \ref{fig:T1F}, \ref{fig:T2F}.

  \begin{figure}[ht]
    \begin{minipage}[b]{0.45\linewidth}
      \centering
      $\xymatrix{
        C_1 \ar@(u,l) \ar@/^/@{.>}[dr] & C_2 \ar@(u,r) \ar@/^/@{.>}[d] & C_5  \ar@(u,r) \ar[d] \ar@/^/@{.>}[dl] \\
        C_3 \ar@(d,l) \ar@/^/@{.>}[r] & C_6 \ar@(r,d) \ar[ul] \ar[u] \ar[ur] \ar[l]  & C_4 \ar@(d,r) \ar[u]
      }$
      \vspace{0.2cm}
      \caption{$\trighyp{F}{1}$}
      \label{fig:T1F}
    \end{minipage}
    \hspace{0.5cm}
    \begin{minipage}[b]{0.45\linewidth}
      \centering
      $\xymatrix{
        C_1 \ar@(u,l) \ar[dr] & C_2 \ar@(u,r) \ar[d] & C_5  \ar@(u,r) \ar[d] \ar[dl] \\
        C_3 \ar@(d,l) \ar[r] & C_6 \ar@(r,d) \ar[ul] \ar[u] \ar[ur] \ar[l]  & C_4 \ar@(d,r) \ar[u]
      }$
      \vspace{0.2cm}
      \caption{$\trighyp{F}{2}$}
      \label{fig:T2F}
    \end{minipage}
  \end{figure}

  To interpret the diagrams:
  \begin{enumerate}
  \item An arrow from a clause $C$ to a clause $D$ represents that $C \in E^k_D$.
  \item A dotted arrow from $C$ to $D$ represents that $\abs{D \sm C} > k$ (so $C \notin E^k_D$), but $C \cap \ol{D} = \es$, and thus for some large enough $k' > k$ we will have $C \in E^{k'}_D$.
  \item No arrow between $C$ and $D$ indicates that $C \cap \ol{D} \not= \es$ (i.e., for all $k'$ we have $C \notin E^k_D$ and $D \notin E^k_C$).
  \item The size of a hyperedge $E^k_D$ is the in-degree of the vertex $D$.
  \end{enumerate}
  Consider $E^1_{C_6} = \set{C_6}$ and $E^2_{C_6} = \set{C_1,C_2,C_3,C_5,C_6}$. As we will see in Lemma \ref{lem:transreptrighyp}, therefore every $F' \sse F$ equivalent to $F$ such that $F' \in \Urefc_1$ must have $C_6 \in F'$. However, $E^2_{C_6}$ contains more clauses than $E^1_{C_6}$, and for example $F \sm \set{C_6} \in \Urefc_2 \sm \Urefc_1$ as shown in Example 8.2 of \cite{GwynneKullmann2012Slur,GwynneKullmann2012SlurJ}. Using the above diagrammatic notation, we can also see that for all $k' \ge 2$ we have $\trighyp{F}{k'} = \trighyp{F}{2}$, as there are no dotted lines for $\trighyp{F}{2}$ (i.e., no clauses $C$ and $D$ such that $\abs{D \sm C} > 2$ but $C \cap \ol{D} = \es$).
\end{examp}

The point of the trigger hypergraph $\trighyp{F}{k}$ is, that every clause-set equivalent to $F$ and of w-hardness at most $k$ must be a transversal of it:
\begin{lem}\label{lem:transreptrighyp}
  Consider $k \in \NNZ$ and $F \in \Cls$ with $\whardness(F) \le k$. Then there is a clause-set $F'$ such that
  \begin{enumerate}
  \item\label{lem:transreptrighyp1} $F' \sse \primec_0(F)$ and $F'$ is equivalent to $F$;
  \item\label{lem:transreptrighyp2} there is an injection $i: F' \ra F$ such that $\fa\, C \in F' : C \sse i(C)$;
  \item\label{lem:transreptrighyp3} $\whardness(F') \le k$;
  \item\label{lem:transreptrighyp4} $F'$ is a transversal of $\trighyp{F}{k}$.
  \end{enumerate}
\end{lem}
\begin{prf}
  Obtain $F'$ from $F$ by choosing for every $C \in F$ some $C' \in \primec_0(F)$ with $C' \sse C$. Then the first two properties are obvious, while Property \ref{lem:transreptrighyp3} follows from Part 1 of Lemma 6.1 in \cite{Ku00g}. Assume that $F'$ is not a transversal of $\trighyp{F}{k}$, that is, there is $C \in \primec_0(F)$ with $F' \cap E^k_C = \es$. Then $\vp_C * F' \in \Usat$, but every clause has length strictly greater than $k$, and thus $k$-resolution does not derive $\bot$ from $\vp_C * F'$, contradicting $\whardness(F') \le k$. \Qed
\end{prf}

Our lower bound method is now captured by the following theorem, which directly follows from Lemma \ref{lem:transreptrighyp}:
\begin{thm}\label{thm:triggersetmethod}
  For $k \in \NNZ$ and $F \in \Wrefc_k$ we have $c(F) \ge \tau(\trighyp{F}{k})$.
\end{thm}
Instead of lower-bounding the transversal number of $\trighyp{F}{k}$, we use that every transversal has to have at least as many elements as there are disjoint hyperedges. So let \bmm{\nu(G)} be the \textbf{matching number} of hypergraph $G$, the maximum number of pairwise disjoint hyperedges; we have $\tau(G) \ge \nu(G)$ for all hypergraphs $G$. So we have to show that there is a set $S \sse \primec_0(F^k_h)$ of exponential size, such that the hyperedges $E^k_C$ for $C \in S$ are pairwise disjoint. For $F^k_h$ we use the doped clause-set $\doping(\smuo(T))$ as considered in Subsection \ref{sec:appSMU1dop}, where the special trees $T$ are constructed in the subsequent subsection.

\subsection{Extremal trees}
\label{sec:est}

For a given hardness $k \ge 1$ we need to construct (full binary) trees which are as large as possible; this is achieved by specifying the height, and using trees which are ``filled up'' completely for the given parameter values:
\begin{defi}\label{def:exstrahltree}
  A pair $(k, h) \in \NNZ^2$ with $h \ge k$ and $k=0 \Ra h=0$ is called an \textbf{allowed parameter pair}. For an allowed parameter pair $(k,h)$ a full binary tree $T$ is called an \textbf{extremal tree of Horton-Strahler number \bmm{k} and height \bmm{h}} if
  \begin{enumerate}
  \item $\hts(T) = k$, $\height(T) = h$;
  \item for all $T'$ with $\hts(T') \le k$ and $\height(T') \le h$ we have $\nds(T') \le \nds(T)$.
  \end{enumerate}
   We denote the set of all extremal trees with Horton-Strahler number $k$ and height $h$ by \bmm{\exhst{k}{h}}.
\end{defi}
Note that for allowed parameter pairs $(k,h)$ we have $k = 0 \Lra h = 0$. Extremal trees are easily characterised and constructed as follows:
\begin{enumerate}
\item $\exhst 00$ contains only the trivial tree (with one node).
\item $\exhst 1h$ for $h \in \NN$ consists exactly of the full binary trees $T$ with $\hts(T) = 1$ and $\height(T) = h$, which can also be characterised as those full binary trees $T$ with $\height(T) = h$ such that every node has at least one child which is a leaf.
\item For $k \ge 2$ and $h \ge k$ we have $T \in \exhst{k}{h}$ iff $T$ has the left subtree $T_0$ and the right subtree $T_1$, and there is $\ve \in \set{0,1}$ with $T_\ve \in \exhst{k-1}{h-1}$ and $T_{1-\ve} \in \exhst{\min(k,h-1)}{h-1}$.
\end{enumerate}
\begin{lem}\label{lem:existextr}
  For all allowed parameter pair $(k,h)$ we have $\exhst{k}{h} \not= \es$.
\end{lem}
The unique elements of $\exhst kk$ for $k \in \NNZ$ are the perfect binary trees of height $k$, which are the smallest binary trees of Horton-Strahler number $k$.

\begin{lem}\label{lem:numexleaves}
 For an allowed parameter pair $(k, h)$ and for $T \in \exhst{k}{h}$ we have $\nlvs(T) = \bmm{\exstrahlersize{k}{h}} := \sum_{i=0}^k \binom{h}{i}$. We have $\alpha(k,h) = \Theta(h^k)$ for fixed $k$.
\end{lem}
\begin{prf}
For $k \le 1$ we have $\alpha(0,0) = 1$ and $\alpha(1,h) = 1 + h$. which are obviously correct. Now consider $k \ge 2$. By induction hypothesis we get
\begin{displaymath}
  \nnds(T) = \alpha(k-1,h-1) + \alpha(\min(k,h-1),h-1).
\end{displaymath}
If $h = k$, then $\alpha(k,h) = 2^k$ (for all $k$), and we get $\nnds(T) = \alpha(k-1,k-1) + \alpha(k-1,k-1) = 2 \cdot 2^{k-1} = 2^k = \alpha(k,k)$. Otherwise we have
\begin{multline*}
  \nnds(T) = \alpha(k-1,h-1) + \alpha(k,h-1) =\\
  \sum_{i=0}^{k-1} \binom{h-1}{i} + \sum_{i=0}^{k} \binom{h-1}{i} = \binom{h-1}{0} + \sum_{i = 1}^k \binom{h-1}{i-1} + \binom{h-1}{i} =\\
             \binom{h-1}{0} + \sum_{i = 1}^k \binom{h}{i} = \sum_{i=0}^{k} \binom{h}{i} = \alpha(k,h).
\end{multline*}
\Qed
\end{prf}

\begin{examp}\label{exp:extt}
  Consider the following labelled binary tree $T$:
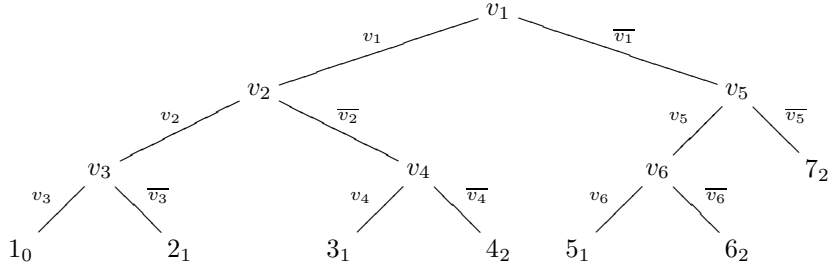
  \begin{displaymath}
    \xygraph{
      !{0;/r7ex/:}
        []{v_1} (
          - [dlll]{v_2}_{v_1} (
            -[dll]{v_3}_{v_2} (
              -[dl]{1_0}_{v_3} (),
              -[dr]{2_1}^{\ol{v_3}} ()
            ),
            -[drr]{v_4}^{\ol{v_2}} (
              -[dl]{3_1}_{v_4} (),
              -[dr]{4_2}^{\ol{v_4}} ()
            )
          ),
          - [drrr]{v_5}^{\ol{v_1}} (
            -[dl]{v_6}_{v_5} (
              -[dl]{5_1}_{v_6} (),
              -[dr]{6_2}^{\ol{v_6}} ()
            ),
            -[dr]{7_2}^{\ol{v_5}} ()
          )
        )
    }
  \end{displaymath}
  Applying the recursive construction/characterisation we see $T \in \exhst{2}{3}$. By simple counting we see that $T$ has $7$ leaves, in agreement with Lemma \ref{lem:numexleaves}, i.e.,  $\sum_{j=0}^2 \binom{3}{j} = \binom{3}{0} + \binom{3}{1} + \binom{3}{2} = 1 + 3 + 3 = 7$. Assuming that of the two subtrees at an inner node, the left subtree has Horton-Strahler numbers as least as big as the right subtree, the idea is that the sum runs over the \emph{number $j$ of right turns in a path from the root to the leaves}. In the above tree $T$, the number of right turns is indicated as an index to the leaf-name. If the Horton-Strahler number is $k$, with at most $k$ right-turns we must be able to reach every leaf.
\end{examp}

We summarise the additional knowledge over Theorem \ref{thm:sumdsmuo} (using additionally that most leaves of $T \in \exhst kh$ have depth precisely $h$):
\begin{lem}\label{lem:sizeex}
  Consider an allowed parameter pair $(k, h)$ and $T \in \exhst kh$, and let $F := \smuo(T)$.
  \begin{enumerate}
  \item $n(\doping(F)) = 2\cdot\alpha(k,h) - 1$ ($ = \Theta(h^k)$ for fixed $k$).
  \item $c(\doping(F)) = \alpha(k,h)$ ($ = \Theta(h^k)$ for fixed $k$).
  \item $\ell(\doping(F)) \le h \cdot \alpha(k,h)$ ($ = \Theta(h^{k+1})$ for fixed $k$).
  \item $\doping(F) \in \Urefc_k \sm \Urefc_{k-1}$ (for $k \ge 1$).
  \end{enumerate}
\end{lem}
In Theorem \ref{thm:separation} we will see that these $\doping(F)$ from Lemma \ref{lem:sizeex} do not have short equivalent clause-sets of hardness $k-1$. A simple example demonstrates the separation between $\Urefc_0$ and $\Urefc_1$ (similar to \cite{Val1994UnitResolutionComplete}, Example 2, which uses Example 6.1 from \cite{KeanTsiknis1990IncrementalPrimeImp}):
\begin{examp}\label{exp:sepUC01}
  The strongest separation is obtained by using $F_h := \doping(\smuo(T))$ for $T \in \exhst 1h$ and $h \in \NN$:
  \begin{enumerate}
  \item $\smuo(T)$, when considering all possible $T$, covers precisely the saturated minimally unsatisfiable renamable Horn clause-set with $h$ variables, which is up to isomorphism equal to $\set{\set{v_1},\set{\ol{v_1},v_2},\dots,\set{\ol{v_1},\dots,\ol{v_{h-1}},v_h}, \set{\ol{v_1},\dots,\ol{v_h}}}$. By Lemma \ref{lem:characsmuhts1} these are precisely those $F \in \Smusati{\delta=1}$ with $n(F) \ge 1$ which contain a full clause.
  \item $n(F_h) = 2h + 1$, $c(F_h) = h+1$, and $\hardness(F_h) = 1$.
  \item $\abs{\primec_0(F_h)} = 2^{h+1}-1$.
  \end{enumerate}
  Considering $G_n := \set{\set{v_1},\dots,\set{v_n},\set{\ol{v_1},\dots,\ol{v_n}}}$ for $n \ge 2$ and $F_n := \doping(G_n)$ we obtain an example similar (but simpler) to Example 6.1 from \cite{KeanTsiknis1990IncrementalPrimeImp}:
  \begin{enumerate}
  \item $n(G_n) = n$ and $c(G_n) = n+1$.
  \item $G_n \in \Musati{\delta=1} \sm \Smusati{\delta=1}$. The above clause-sets $\smuo(T)$ are obtained precisely as saturations of the $G_n$ (due to Lemma \ref{lem:characsmuhts1}; a saturation adds literal occurrences until we obtain a saturated minimally unsatisfiable clause-set).
  \item $\mps(G_n)$ consists precisely of the subsets of $G_n$ containing the negative clause, plus the singleton-subsets given by the unit-clauses.
  \item Thus $\abs{\mps(G_n)} = 2^n + n$.
  \item $n(F_n) = 2n+1$, $c(F_n) = n+1$, and $\hardness(F_n) = 1$.
  \item $\abs{\primec_0(F_n)} = 2^n + n$.
  \end{enumerate}
\end{examp}

\subsection{The exponential lower bound}
\label{sec:hierbadrep}

The task is to find many disjoint hyperedges in $\trighyp{F^k_h}{k}$, where $F^k_h := \doping(\smuo(T))$ for $T \in \exhst {k+1}h$. Our method for this is to show that there are many ``incomparable'' subsets of leaves in $T$ in the following sense. The \emph{depth} of a node $w$ in a rooted tree $T$, denoted by $\bmm{\depth_T(w)} \in \NNZ$, is the length of the path from the root of $T$ to $w$. Recall that two sets $A, B$ are \emph{incomparable} iff $A \not\sse B$ and $B \not\sse A$. Furthermore we call two sets $A, B$ \emph{incomparable on a set $C$} if the sets $A \cap C$ and $B \cap C$ are incomparable.
\begin{defi}\label{def:depthkincomp}
  Consider a full binary tree $T$, where every leaf has depth at least $k+1$. Consider furthermore $\es \subset V, V' \sse \lvs(T)$. Then $V$ and $V'$ are \textbf{depth-\bmm{k}-incomparable for \bmm{T}} if $V$ and $V'$ are incomparable on $\lvs(T_w)$ for all $w \in \nds(T)$ with $\depth_T(w) = k$.
\end{defi}
Note that for all allowed parameter pairs $(k,h)$ and $T \in \exhst kh$ every leaf has depth at least $k$.

\begin{lem}\label{lem:depthksephyperedge}
  Consider $k \in \NNZ$, $T \in \Tsmuo$, and $\es \not= V_0, V_1 \sse \lvs(T)$ which are depth-$k$-incomparable for $T$. Let $F := \smuo(T)$ and consider $\trighyp Fk$ (recall Definition \ref{def:triggerhypergraph}).  Then the hyperedges $E^k_{C_{V_0}}$, $E^k_{C_{V_1}}$ are disjoint (recall Definition \ref{def:clausesDSMU1}).
\end{lem}
\begin{prf}
Assume that $E^k_{C_{V_0}}$, $E^k_{C_{V_1}}$ are not disjoint; thus there is $\es \not= V \sse \lvs(T)$ with $C_V \in E^k_{C_{V_0}} \cap E^k_{C_{V_1}}$. We will show that there is $\ve \in \set{0,1}$ with $\abs{C_V \sm C_{V_\ve}} \ge k+1$, which contradicts the definition of $\trighyp Fk$.

Since $V \not= \es$, there is $w \in V$. Consider the first $k+1$ nodes $w_1,\dots,w_{k+1}$ on the path from the root to $w$. Let $w_i'$ be the child of $w_{i-1}$ different from $w_i$ for $i \in \tb 2{k+1}$, and let $T_i := T_{w_{i+1}'}$ for $i \in \tb 1k$, while $T_{k+1} := T_{w_{k+1}}$; see Figure \ref{fig:depthksep}. We show that each of $T_1, \dots, T_{k+1}$ contributes at least two unique literals to $\abs{C_V \sm C_{V_0}} + \abs{C_V \sm C_{V_1}}$, so that we get $\abs{C_V \sm C_{V_0}} + \abs{C_V \sm C_{V_1}} \ge (k+1) \cdot 2$, from which follows that there is $\ve \in \set{0,1}$ with $\abs{C_V \sm C_{V_\ve}} \ge k+1$ as claimed.

\begin{figure}[h]
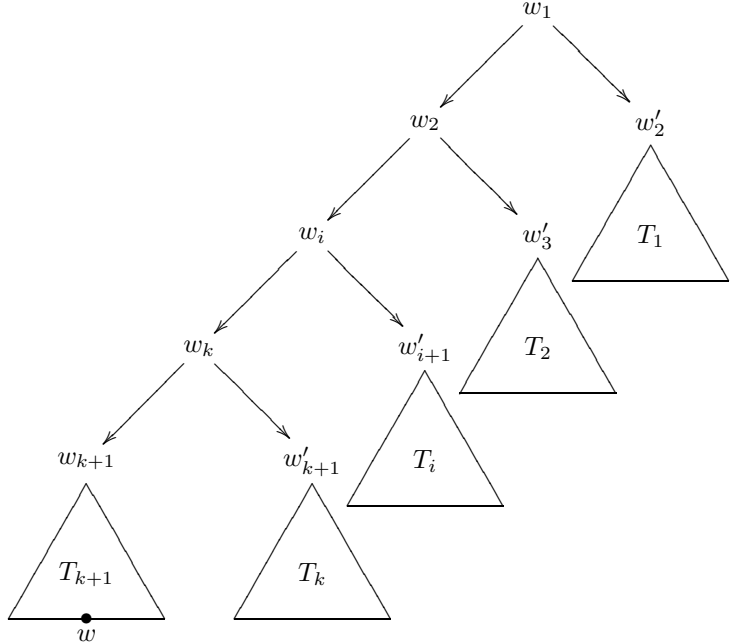

  \[\xygraph{
    !{(-4,-5.5 )}*+{}
    !{0;/r1.5cm/:}
    !{(0,0) }*+{w_1}="w1"
    !{(-1,-1) }*+{w_2}="w2"
    !{(1,-1) }*+{w_2'}="w1p"
    !{(-2,-2) }*+{w_i}="wi"
    !{(0,-2) }*+{w_3'}="w2p"
    !{(-3,-3)}*+{w_k}="wk"
    !{(-1,-3) }*+{w_{i+1}'}="wip"
    !{(-4,-4) }*+{w_{k+1}}="wkpp"
    !{(-2,-4) }*+{w_{k+1}'}="wkp"
    !{(-2,-2.5) }*+{\xypolygon3"tkpd"{~:{(0.8,0):}}}="wkpptree"
    !{(-1,-2.5) }*+{\xypolygon3"tkd"{~:{(0.8,0):}}}="wkptree"
    !{(-0.5,-2.0) }*+{\xypolygon3"tid"{~:{(0.8,0):}}}="wiptree"
    !{(0,-1.5) }*+{\xypolygon3"t2d"{~:{(0.8,0):}}}="w2ptree"
    !{(0.5,-1) }*+{\xypolygon3"t1d"{~:{(0.8,0):}}}="w1ptree"
    !{(-4,-5) }*+{T_{k+1}}="tkp"
    !{(-2,-5) }*+{T_k}="tk"
    !{(-1,-4) }*+{T_i}="ti"
    !{(0,-3) }*+{T_2}="t2"
    !{(1,-2) }*+{T_1}="t1"
    !{(-4,-5.4)}*+{\bullet}="wb"
    !{(-4,-5.55)}*+{w}="w"
    "w1" : "w2"
    "w1" : "w1p"
    "w2" : "wi"
    "w2" : "w2p"
    "wi" : "wk"
    "wi" : "wip"
    "wk" : "wkpp"
    "wk" : "wkp"
    }\]
  \caption{Illustration of sub-trees $T_1,\dots,T_{k+1}$.}
  \label{fig:depthksep}
\end{figure}

Due to the depth-k-incomparability of $V, V'$, for each $i \in \tb 1{k+1}$ and each $\ve \in \set{0,1}$ there are nodes $v_i^{\ve}$ with $v_i^{\ve} \in (\lvs(T_i) \cap V_{\ve}) \sm V_{\ol{\ve}}$. We have two cases now:
\begin{enumerate}
\item[I] If $v_i^{\ve} \in V$, then $u_{v_i^{\ve}} \in C_V \sm C_{V_{\ol{\ve}}}$.
\item[II] If $v_i^{\ve} \notin V$, then consider the first node $v$ on the path from $v_i^{\ve}$ to the root such that for the other child $v'$ of $v$, not on that path to the root, holds $\lvs(T_{v'}) \cap V \not= \es$: now for the literal $x$ labelling the edge from $v$ to $v'$ we have $x \in C_V \sm C_{V_{\ve}}$. Note that $v$ is below or equal to $w_i$ (due to $w \in V$).
\end{enumerate}
For each $\ve \in \set{0,1}$, the literals collected in $C_V \sm C_{V_{\ve}}$ from these $k+1$ sources do not coincide, due to the pairwise node-disjointness of the trees $T_1, \dots, T_{k+1}$. \Qed
\end{prf}

\begin{thm}\label{thm:nogoodksoft}
  Consider $k \in \NNZ$, $h \ge k+1$, and $T \in \exhst {k+1}h$; let $F := \doping(\smuo(T))$ and $m := \exstrahlersize{1}{h-k} = 1 + h - k$. We have
  \begin{displaymath}
    \nu(\trighyp{F}{k}) \ge \binom{m}{\floor{\frac{m}{2}}} > \frac{1}{\sqrt{2}} \frac{2^m}{\sqrt{m}} = \Theta(\frac{2^h}{\sqrt{h}}),
  \end{displaymath}
  where the second inequality assumes $h \ge k+5$, while the $\Theta$-estimation assumes fixed $k$.
\end{thm}
\begin{prf}
  For every $S \sse \pot(\lvs(T))$ with $\es \notin S$, such that every two different elements of $S$ are depth-$k$-incomparable for $T$, we have $\nu(\trighyp{F}{k}) \ge \abs{S}$ by Lemma \ref{lem:depthksephyperedge}. We can actually determine the maximal size of such an $S$, which is $M := \binom{m}{m'}$, where $m' := \floor{\frac{m}{2}}$, as follows. Let $\TT := \set{T_w : w \in \nds(T) \und \depth_T(w) = k}$; note that for $T', T'' \in \TT$ with $T' \not= T''$ we have $\lvs(T') \cap \lvs(T'') = \es$. Choose $T_0 \in \TT$ with minimal $\nlvs(T_0)$; by Lemma \ref{lem:numexleaves} we have $\nlvs(T_0) = m$. Let $S_0 := \set{V \cap \lvs(T_0) : V \in S}$. Then $S_0$ is an antichain (i.e., the elements of $S_0$ are pairwise incomparable) and $\abs{S_0} = \abs{S}$. By Sperner's Theorem (\cite{Sperner1928SubsetsFiniteSets}) holds $\abs{S_0} \le M$, and this upper bound $M$ is realised, just observing the antichain-condition, by choosing for $S_0$ the set $\binom{\lvs(T_0)}{m'}$ of subsets of $\lvs(T_0)$ of size $m'$. This construction of $S_0$ can be extended to a construction of $S$ (of the same size) by choosing for each $T' \in \TT$ an injection $j_{T'}: S_0 \ra \binom{\lvs(T')}{m'}$ and defining $S := \set{\bc_{T' \in \TT} j_{T'}(V)}_{V \in S_0}$. The given estimation of $M$ follows from Stirling's approximation. \Qed
\end{prf}

We are now able to state the main result of this article, proving Conjecture 1.1 from \cite{GwynneKullmann2012Slur,GwynneKullmann2012SlurJ} that $\Urefc_k$, and indeed also $\Wrefc_k$, is a proper hierarchy of boolean functions regarding polysize representations without auxiliary variables:
\begin{thm}\label{thm:separation}
  Consider $k \in \NNZ$. We have $\Wrefc_k \nosim \Urefc_{k+1}$. The details are as follows.

  For $h \ge k+1$ choose one $T_h \in \exhst {k+1}h$ (note there is up to left-right swaps exactly one element in $\exhst {k+1}h$), and let $F_h := \doping(\smuo(T_h))$. Consider the sequence $(F_h)_{h \ge k+1}$.
  \begin{enumerate}
  \item By Lemma \ref{lem:sizeex} we have $n(F_h) = \Theta(h^{k+1})$ as well as $c(F_h) = \Theta(h^{k+1})$, and $F_h \in \Urefc_{k+1}$.
  \item Consider a sequence $(F_h')_{h \ge k+1}$ of clause-sets with $F_h'$ equivalent to $F_h$, such that $F_h' \in \Wrefc_k$. By Theorems \ref{thm:nogoodksoft}, \ref{thm:triggersetmethod} we have $c(F_h') = \Omega(\frac{2^h}{\sqrt{h}})$.
  \end{enumerate}
\end{thm}

We conjecture that Theorem \ref{thm:separation} can be strengthened by including the PC-hierarchy in the following way:
\begin{conj}\label{con:sepsharp}
  For every $k \in \NNZ$ we have $\Wrefc_k \nosim \Propc_{k+1}$.
\end{conj}
A step towards Conjecture \ref{con:sepsharp} is Theorem \ref{thm:sepUCPC}.

\section{Knowledge compilation properties}
\label{sec:hierkcprop}

In view of the above separation result for the hierarchies $\Urefc_k, \Propc_k$ and $\Wrefc_k$, we now place these hierarchies in their context in the knowledge compilation (KC) literature. First we need to review the basic setting. A boolean function $f$ is considered here as a map $f: \Tass(V) \ra \set{0,1}$, where $V \subset \Va$ is a finite set of variables, while $\Tass(V) := \set{\vp \in \Pass : \var(\vp) = V}$ is the set of all total assignments on $V$. For every $\vp \in \Pass$ with $\var(\vp) \supseteq V$ the value $f(\vp) \in \set{0,1}$ is defined via restriction of $\vp$ to $V$. Knowledge compilation is about ``representations'' $F$ (in a general sense) of boolean functions $f$ such that basic queries can be answered efficiently:
\begin{itemize}
\item \ul{Co}nsistency checking (\textrm{CO}): input $F$, whether $f$ is constant $0$ or not.
\item \ul{C}lausal \ul{e}ntailment checking (\textrm{CE}): input $F$ and clause $C$, whether $\vp * \set{C} = \top$ for all $\vp \in \Tass(V)$ with $f(\vp) = 1$.
\item \ul{Va}lidity checking (\textrm{VA}): input $F$, whether $f$ is constant $1$ or not.
\item \ul{Im}plicant checking (\textrm{IM}): input $F$ and partial assignment $\vp$ with $\var(\vp) \sse V$, whether $f(\vp') = 1$ for all $\vp' \in \Tass(V)$ with $\vp' \supseteq \vp$.
\item \ul{S}emantic \ul{E}ntailment (\textrm{SE}): input $F, F'$, whether $f \models f'$ (i.e., whether for all $\vp$ with $\var(f) \cup \var(f') \sse \var(\vp)$ holds $f(\vp) = 1 \Ra f'(\vp) = 1$).
\item \ul{Eq}uivalence checking (\textrm{EQ}): input $F, F'$; whether $f \models f'$ and $f' \models f$.
\item \ul{M}odel \ul{E}numeration (\textrm{ME}): input $F$, enumerate $\vp \in \Tass(V)$ with $f(\vp) = 1$.
\item \ul{M}odel \ul{C}ounting (\textrm{MC}): input $F$, count $\vp \in \Tass(V)$ with $f(\vp) = 1$.
\end{itemize}
The motivation of \cite{Val1994UnitResolutionComplete} for defining $\Urefc$ was to introduce a class of clause-sets for knowledge compilation, such that these basic queries have the same query complexity as the PI class (where PI is the same as $\Urefc_0$). Theorem \ref{thm:kcprop} now shows that $\Urefc_k$, $\Propc_k$ and $\Wrefc_k$ all fulfil the same criteria. This result along with Theorem \ref{thm:separation} means, that $\Urefc_k$, $\Propc_k$ and $\Wrefc_k$ offer intermediate target classes for knowledge compilation inbetween the CNF and PI classes, where the parameter $k$ allows query time to be traded for size. Every fixed level of each hierarchy (except of $\Propc_0$) is a complete class with respect to representation of boolean functions, unlike classes such as $\Pcls{2}$ (CNF clause-sets with clauses of size at most two) or $\Ho$ (Horn clause-sets), or other hierarchies for polynomial time satisfiability like $\Rho_k$ (generalised renamable Horn clause-sets), each of which is included (as classes) at some fixed level of $\Urefc_k$.

\begin{thm}\label{thm:kcprop}
  For all fixed $k \in \NNZ$ and all $F,F' \in \Wrefc_k \spe \Urefc_k \spe \Propc_k$, the queries CO, CE, VA, IM, EQ, SE (as specified above) are decidable in polynomial time (in $\ell(F)$). Furthermore, \textrm{ME}, i.e., enumerating all satisfying assignments, is possible in time $p(\ell(F),m)$ for some fixed polynomial $p$, where $m$ is the number of satisfying total assignments for $F$.
\end{thm}
\begin{prf}
  That clausal entailment is decidable in poly-time for $\Wrefc_k$ is shown in Subsection 6.5 of \cite{Ku99b}. Since we are dealing with clause-sets (conjunctions of clauses), this implies that the other five query-decisions can be done in polynomial time (where $F$ is valid (a tautology) iff $F = \top$, while $\vp$ corresponds to an implicant iff $\vp * F = \top$). Finally, that all models can be enumerated in poly-time in $\ell(F)$ and $m$ follows from the fact that we can build a decision tree with at most $m$ true-leaves and at most $n(F) \cdot m$ false-leaves (compare Lemma A.3 in \cite{DarwicheMarquis2002KCmap}). \Qed
\end{prf}

We finish with an overview on the status of polytime queries for our classes and well-known KC classes in Figure \ref{fig:introkcclassqueries}:
\begin{itemize}
\item NNF means ``negation normal form'', which are circuits with AND's and OR's or unbounded fan-in and where the inputs are literals.
\item DNNF means ``decomposable NNF'', that is, different children of any AND do not have common variables.
\item d-DNNF means ``deterministic DNNF'', where additionally any two different children of any OR must be logically contradictory.
\item DNF's (disjunctive normal forms) are special DNNF's.
  \begin{itemize}
  \item DNF's are just clause-sets, but now interpreted as disjunction of conjunction (not as CNF's, which is the default for clause-sets, that is, conjunctions of disjunctions).
  \item The intersection of d-DNNF and DNF is the class of orthogonal DNF's (as clause-sets the hitting clause-sets, that is, each two different clauses have at least one clash).
  \end{itemize}
\item MODS (like ``models'') are the DNF's where each DNF-clause contains all variables (so these are special orthogonal DNF's).
\item IP means prime implicants, which as clause-sets is the same is PI (prime implicates), that is, $\Urefc_0 = \Wrefc_0$ after removal of subsumed clauses, but interpreted as DNF's.
\item BDD means ``binary decision diagrams'', OBDD means ``ordered (reduced) BDD'', while for OBDD$_{\le}$ one global order on the variables is used.
\end{itemize}

\begin{figure}[H]
\newcommand{\thickline}{\Xhline{3\arrayrulewidth}}
  \centering
  \begin{tabular}{|c|c|c|c|c|c|c|c|c|}
    \hline
    \textbf{L} & \textbf{CO} & \textbf{VA} & \textbf{CE} & \textbf{IM} & \textbf{EQ} & \textbf{SE} & \textbf{CT} & \textbf{ME} \\ \hline\hline
    \textrm{NNF} & $\circ$ & $\circ$ & $\circ$ & $\circ$ & $\circ$ & $\circ$ & $\circ$ & $\circ$ \\ \hline
    \textrm{DNNF} & \tick & $\circ$ & \tick & $\circ$ & $\circ$ & $\circ$ & $\circ$ & \tick \\ \hline
    \textrm{d-DNNF} & \tick & \tick & \tick & \tick & ? & $\circ$ & \tick & \tick \\ \thickline\
    \textrm{BDD} & $\circ$ & $\circ$ & $\circ$ & $\circ$ & $\circ$ & $\circ$ & $\circ$ & $\circ$ \\ \hline
    \textrm{OBDD} & \tick & \tick & \tick & \tick & \tick & $\circ$ & \tick & \tick \\ \hline
    \textrm{OBDD$_{\le}$} & \tick & \tick & \tick & \tick & \tick & \tick & \tick & \tick \\ \thickline
    \textrm{DNF} & \tick & $\circ$ & \tick & $\circ$ & $\circ$ & $\circ$ & $\circ$ & \tick \\ \hline
    \textrm{IP} & \tick & \tick & \tick & \tick & \tick & \tick & $\circ$ & \tick \\ \hline
    \textrm{MODS} & \tick & \tick & \tick & \tick & \tick & \tick & \tick & \tick \\ \thickline
    \textrm{CNF} & $\circ$ & \tick & $\circ$ & \tick & $\circ$ & $\circ$ & $\circ$ & $\circ$ \\ \hline
   \bmm{\Wrefc_k} & \textbf{\tick} & \textbf{\tick} & \textbf{\tick} & \textbf{\tick} & \textbf{\tick} & \textbf{\tick} & \bmm{\circ} & \textbf{\tick} \\ \hline
    \bmm{\Urefc_k} & \textbf{\tick} & \textbf{\tick} & \textbf{\tick} & \textbf{\tick} & \textbf{\tick} & \textbf{\tick} & \bmm{\circ} & \textbf{\tick}\\ \hline
  \end{tabular}
  \caption{Subsets of the \textrm{NNF} language and their corresponding polytime queries. \tick means ``query possible in polytime'', $\circ$ means ``not possible in polytime (in general) unless P = NP'', and ? means that there is no known result either way. Results for $\Wrefc_k, \Urefc_k$ are from Theorem \ref{thm:kcprop}, and all other results are from \cite{DarwicheMarquis2002KCmap}.}
  \label{fig:introkcclassqueries}
\end{figure}

\section{Separating $\Urefc$ from $\Propc$}
\label{sec:sepUCPC}

Using \cite{BBCGKV2013Propc}, we show now that there is a polysize sequence in $\Urefc$ such that no equivalent polysize sequence exists in $\Propc$ (where the separation in fact is exponential).

\begin{thm}\label{thm:sepUCPC}
  We have $\Propc \nosim \Urefc$. This separation is achieved by the polysize family $(M_q)_{q \in \NN}$ from Theorem 5.8 in \cite{BBCGKV2013Propc}:
  \begin{itemize}
  \item In Definition 5.2 in \cite{BBCGKV2013Propc} for $q \in \NN$, disjoint sets $X, Y, Z$ of size $q$ and $W \sse X \times Y \times Z$ the clause-set $\vp_W$ is defined.
  \item Now $M_q := \vp_{X \times Y \times Z}$ for Theorem 5.8 in \cite{BBCGKV2013Propc}.
  \end{itemize}
  It is shown in \cite{BBCGKV2013Propc} that $(M_q)_{q \in \NN}$ has no equivalent polysize sequence in $\Propc$, while we have that $(M_q)_{q \in \NN}$ is in $\Urefc$.
\end{thm}
\begin{prf}
It remains to show that $M_q \in \Urefc$. The proof sketch is as follows. The ``hidden clause-set'' in $M_q$ is
\begin{displaymath}
  F_q := \set{\set{a_1,\ol{a_2}},\set{a_2,\ol{a_3}},\dots,\set{a_{q-1},\ol{a_q}},\set{a_q,a_1}} \in \Pcls{2} \cap \Sat.
\end{displaymath}
We have $\hardness(F_q) = 1$ (by Lemma 6.6 in \cite{GwynneKullmann2012SlurJ}), and thus also for all partial assignments $\vp$ holds $\hardness(\vp * F_q) \le 1$.\footnote{Furthermore $\phardness(F_q) = 2$ (since the literal $a_1$ is forced), which causes $M_q \notin \Propc$, but this is not our concern here.} Now the clauses of $\vp * F_q$ are the only clauses potentially usable in resolution refutations of $\vp * M_q$, using the fundamental insight from \cite{Ku00f} (or see \cite{Kullmann2007HandbuchMU}), that clauses satisfiable by some autarky can not participate in any resolution refutation, while the definition of $M_q$ precisely makes all clauses in $\vp * M_q$ satisfiable by some autarky, which still contain one of the other variables $b_i^j, c_i^j$. \Qed
\end{prf}

Generalising Theorem \ref{thm:sepUCPC} to a separation of $\Urefc_k$ from $\Propc_k$ for all $k \ge 1$ requires more work:
\begin{conj}\label{con:pcuc}
  For all $k \in \NNZ$ we have $\Propc_k \nosim \Urefc_k$.
\end{conj}
With Theorem \ref{thm:sepUCPC} we know Conjecture \ref{con:pcuc} for $k \le 1$.

\section{Conclusion and open problems}
\label{sec:open}

We conclude by directions for future research.

\subsection{A complete picture}
\label{sec:concfsep}

Conjecture \ref{con:noauxstr} (recall the discussion in Subsection \ref{sec:intromap}) paints a complete picture regarding the relations of the classes $\Propc_k, \Urefc_k, \Wrefc_k$ w.r.t.\ polysize representation of equivalent boolean functions. By Theorem \ref{thm:separation} we have $\Wrefc_k \nosim \Urefc_{k+1}$ for $k \ge 0$, and by Theorem \ref{thm:sepUCPC} we have $\Propc_1 \nosim \Urefc_1$. We now discuss what remains to be shown.

Conjecture \ref{con:sepsharp} claims $\Wrefc_k \nosim \Propc_{k+1}$. This would also imply $\Propc_k \nosim \Propc_{k+1}$ (we already know $\Urefc_k \nosim \Urefc_{k+1}$ and $\Wrefc_k \nosim \Wrefc_{k+1}$, while currently we only know $\Propc_k \nosim \Propc_{k+2}$).

Conjecture \ref{con:pcuc} claims $\Propc_k \nosim \Urefc_k$, and additionally we have
\begin{conj}\label{con:UCWCnoaux}
  For all $k \ge 2$ holds $\Urefc_k \nosim \Wrefc_k$.
\end{conj}

Since $\Wrefc_3 \not\sse \Urefc_k$, we get
\begin{conj}\label{con:UCWCnoauxstr}
  For all $k \ge 0$ holds $\Urefc_k \nosim \Wrefc_3$.
\end{conj}

And from $\Wrefc_2 \not\sse \Urefc_3$ we get
\begin{conj}\label{con:UCWCnoauxstr2a}
  $\Urefc_3 \nosim \Wrefc_2$.
\end{conj}

Conjectures \ref{con:UCWCnoauxstr}, \ref{con:UCWCnoauxstr2a} together imply Conjecture \ref{con:UCWCnoaux}. If Conjecture \ref{con:diffw2u2} is true, then we get the stronger form (which implies Conjectures \ref{con:UCWCnoaux}, \ref{con:UCWCnoauxstr}, \ref{con:UCWCnoauxstr2a}):
\begin{conj}\label{con:UCWCnoauxstr2b}
  For all $k \ge 0$ holds $\Urefc_k \nosim \Wrefc_2$.
\end{conj}

\subsection{Alternative hierarchies for representations}
\label{sec:conclalth}

In a sense, the $\Urefc_k = \Slur_k$ hierarchy unified the three predecessor hierarchies $\Altsluri{k}$ introduced in \cite{Vlcek2010ClassesBoolPolySLUR}, $\Altslurstari{k}$ introduced in \cite{CepekKuceraVlcek2012SLUR}, and $\Canoni{k}$ introduced in \cite{BalyoGurskyKuceraVlcek2012SLURHier}. In \cite{GwynneKullmann2012SlurSOFSEM,GwynneKullmann2012Slur,GwynneKullmann2012SlurJ} we compared them directly (as sets of clause-sets) to the UC-hierarchy, and showed that they were properly included. An interesting question is now whether these more ``shallow'' hierarchies (for each level the CE-queries can be answered in linear time) also form strict hierarchies w.r.t.\ polysize representations without new variables. It is rather easy to see that the hierarchy $\Canoni{k}$ collapses to $\Canoni{0} = \Urefc_0$:
\begin{lem}\label{lem:collapsecanon}
  For $F \in \Cls$ let $k(F)$ be the minimal $k \in \NNZ$ such that $F \in \Canoni{k}$. Then the function $\primec_0: \Cls \ra \Canoni{0} = \Urefc_0$ can be computed in time $O(c(F)^{3 \cdot 2^{k}} \cdot \ell(F))$, when the input is $F$ together with $k := k(F)$.
\end{lem}
\begin{prf}
  Let $K := 2^k$. So for every $C \in \primec_0(F)$ there exists $F' \sse F$ with $F' \models C$ and $c(F') \le K$, since a resolution tree of height $k$ has at most $K$ leaves. Now we compute $\primec_0(F)$ as follows:
  \begin{enumerate}
  \item Set $P := \es$.
  \item Run through all $F' \sse F$ with $c(F') \le K$; their number is $O(c(F)^K)$.
  \item For each $F'$ determine whether $F' \models \purec(F')$ holds, in which case clause $\purec(F')$ is added to $P$; note that the test can be performed in time $O(2^K \cdot K)$.
  \item The final $P$ obtained has $O(c(F)^K)$ many elements. After performing subsumption elimination (in cubic time) we obtain $\primec_0(F)$ (by Lemma \ref{lem:uniquepurec}). \Qed
  \end{enumerate}
\end{prf}
It is an interesting question whether also the hierarchies $\Altsluri{k}$, $\Altslurstari{k}$ collapse or not, and whether they can be reduced to some $\Urefc_k$ (for some fixed $k$).

\subsection{Compilation procedures}
\label{sec:conclcompil}

For a given boolean function $f$ and $k \in \NNZ$, how do we find algorithmically a ``small'' equivalent $F \in \Urefc_k$ ? In \cite{GwynneKullmann2012SlurJ}, Section 8, the notion of a ``$k$-base for $f$'' is introduced, which is an $F \in \Urefc_k$ equivalent to $f$, with $F \sse \primec_0(f)$ and where no clause can be removed without increasing the hardness or destroying equivalence. It is shown that if $f$ is given as a 2-CNF, then a smallest $k$-base is computable in polynomial time, but even for $f$ with given $\primec_0(f)$, where $\primec_0(f)$ is a Horn clause-set, deciding whether a $k$-base of a described size for a fixed $k \ge 1$ exists is NP-complete.

There are interesting applications where $\primec_0(f)$ is given (or can be computed), and where then some small equivalent $F \in \Urefc_k$ is sought. The most basic approach filters out unneeded prime implicates; see \cite{GwynneKullmann2011TranslationsPrelim,GwynneKullmann2011HardnessPrelim} for some initial applications to cryptanalysis. A simple filtering heuristic, used in \cite{GwynneKullmann2011TranslationsPrelim,GwynneKullmann2011HardnessPrelim}, is to favour (keeping) short-clauses. In a first phase, starting with the necessary elements of $\primec_0(f)$, further elements are added (when needed) in ascending order of size for building up the initial $F \in \Urefc_k$ (which in general is not a base). In the second phase, clauses from $F$ are removed in descending order of size when reducing to a $k$-base. The intuition behind this heuristic is that small clauses cover more total assignments (so fewer are needed), and they are also more likely to trigger $\rk_k$, making them more useful in producing small, powerful representations. Essentially the same heuristic is considered in \cite{BordeauxMarquesSilva2012KnowledgeCompilation} (called ``length-increasing iterative empowerment'') when generating representations in $\Propc$.

For the case that $f$ is given by a CNF $F_0$, in \cite{Val1994UnitResolutionComplete,Sinz2002Compilation} one finds refinements of the resolution procedure applied to $F_0$, which would normally compute $\primec_0(f)$, i.e., the $0$-base in $\Urefc_0$, and where by some form of ``compression'' now an equivalent $F \in \Urefc_1$ is computed. This approach needed to be generalised to arbitrary $\Urefc_k$.

\subsection{Allowing auxiliary variables}
\label{sec:conclstrict}

In this report we considered representations of boolean functions by equivalent CNF-clause-sets, to be used in constructing ``good'' SAT translations or in knowledge compilation (KC). An important advantage of this approach is the ability to systematically search for good representations, as discussed in the previous Subsection \ref{sec:conclcompil}. However it is also well-known that without the use of auxiliary variables, many relevant boolean functions do only have very large equivalent CNF-clause-sets at all. So we consider now the extension of the picture, as developed in the previous Subsection \ref{sec:concfsep}, by allowing auxiliary variables.

\subsubsection{The notion of ``CNF-representation''}
\label{sec:cnfrep}

First a framework for the meaning of auxiliary variables is needed. In general it is understood that existential quantification of the auxiliary variables is the right condition. In the SAT-context, it seems best to keep the quantification implicit, and we arrive at the following notion: A \textbf{CNF-representation} (possibly with auxiliary variables) of a boolean function $f$ is a clause-set $F$ with $\var(f) \sse \var(F)$ such that the satisfying assignments of $F$, projected to $\var(f)$, are precisely the satisfying assignments of $f$, or, in other words, if for $\vp \in \Tass(\var(f))$ holds $f(\vp) = 1 \Lra \vp * F \in \Sat$.\footnote{To be completely precise, we needed to use ``formal clause-sets'' here, which can have variables actually not occurring in the clauses.} Note that if for a CNF-representation $F$ of $f$ holds $\var(F) = \var(f)$, then $F$ is logically equivalent to $f$. A sequence $(F_n')_{n \in \NN}$ is called a \textbf{CNF-representation of $(F_n)_{n \in \NN}$} if for all $n \in \NN$ the clause-set $F_n'$ is a CNF-representation of $F_n$.

\begin{lem}\label{lem:characrep}
  A clause-set $F \in \Cls$ is a CNF-representation of a boolean function $f$ with $\var(f) \sse \var(F)$ if and only if $\primec_0(f) = \set{C \in \primec_0(F) : \var(C) \sse \var(F)}$.
\end{lem}
\begin{prf}
  Let $V := \var(f)$.
  First assume that $F$ is a CNF-representation of $f$. If $C \in \primec_0(f)$, then due to $F \models f$ we have $F \models C$, and if $C \notin \primec_0(F)$, then there would be $C' \in \primec_0(F)$ with $C' \subset C$, and then for $\vp := \vp_{C'}$ there would be an extension $\vp' \in \Tass(V)$ with $f(\vp') = 1$, but $\vp * F \in \Usat$. For the other inclusion consider $C \in \primec_0(F)$ with $\var(C) \sse V$. If there would be an assignment $\vp \in \Tass(V)$ with $f(\vp) = 1$ but $\vp * \set{C} = \set{\bot}$, then we had $\vp * F \in \Usat$ contradicting $f(\vp) = 1$. So $f \models C$, and due to $F \models f$ it follows $C \in \primec_0(f)$.

  Now assume $\primec_0(f) = \set{C \in \primec_0(F) : \var(C) \sse V}$, and we have to show that $F$ is a CNF-representation of $f$. So first consider $\vp \in \Tass(V)$ with $\vp * F = \top$. If $f(\vp \rstr V) = 0$, then there would be $C \in \primec_0(f)$ with $\vp_C \sse \vp$, but then $C \in \primec_0(F)$, and thus $\vp * F \in \Usat$; so by contradiction $f(\vp \rstr V) = 1$. And if for $\vp \in \Tass(V)$ holds $f(\vp) = 1$ but $\vp * F \in \Usat$, then there is $C \in \primec_0(F)$ with $\vp_C \sse \vp$, and we had $C \in \primec_0(f)$. \Qed
\end{prf}

As a KC ``formalism'' the CNF-representation of boolean functions is known as ``$\ex\mr{CNF}$'', which we write as $\ex\Cls$, defined as the set of pairs $(V,F)$ with $F \in \Cls$ and $V \sse \var(F)$; the variables of $V$ are existentially quantified, and the boolean function represented by $(V,F)$ is given by the QBF $\ex v_1,\dots,\ex v_m : F$, where $V = \set{v_1,\dots,v_m}$. Evaluation of the underlying boolean function is an NP-complete task, and so restrictions are needed to obtain efficient representations. A natural restriction is to demand that evaluation can be done by unit-clause propagation, and it is well-known that via the Tseitin-translation this corresponds, modulo linear-time transformations, to the circuit-representation of boolean functions (see \cite{BubeckBuening2010Definitions} for closely related results). We call that class $\ex\Up$, the class of $(V,F)$ with $F \in \Cls$, $V \sse \var(F)$, such that for all $\vp \in \Tass(\var(F) \sm V)$ holds $\rk_1(\vp * F) \in \set{\top, \set{\bot}}$. This class is what people think of first (intuitively) when they have to represent a boolean function $f$ via CNF: first some polynomial time mechanism for computing $f$ is sought, then this is translated into a boolean circuit, which via the Tseitin-translation is translated in $\ex\Up$. Using $V = \es$, the class $\Cls$ is trivially simulated by $\ex\Up$, while it is easy to come up with examples in $\ex\Up$ which have no polysize representations in $\Cls$ (which means, in the KC context, CNF-representations without new variables; see for example Lemma \ref{lem:xorabsc}).

\subsubsection{Absolute and relative condition}
\label{sec:absrel}

The question then is how to treat the classes $\Propc_k, \Urefc_k, \Wrefc_k$ in this context. There are two possibilities, namely that the conditions constituting these classes also concern the auxiliary variables or not. The first case, that for a CNF representation $F$ of $f$ we simply require $F \in \Propc_k, \Urefc_k, \Wrefc_k$ resp., we call the \emph{absolute condition}, while the second case we call the \emph{relative condition}. In the context of KC, the class $\Urefc$ under the absolute condition is denoted by $\UrefcE$ (\cite{DarwicheMarquis2002KCmap}), while $\EUrefc$ denotes the usage of the relative condition (\cite{FargierMarquis2008Closure}).

$\UrefcE$ can be defined as pairs $(V,F)$, where $F \in \Urefc$, while $V \sse \var(F)$ are the variables which are existentially quantified (the auxiliary variables). Similarly we get the ``formalisms'' $\Urefc_k[\ex], \Propc_k[\ex], \Wrefc_k[\ex]$ for KC; the boolean functions represented by $(V,F)$ are obtained from the boolean functions given by the CNF $F$ by projecting the satisfying assignments to $V$. However $\EUrefc$ can not be defined just from the class $\Urefc$, but the underlying condition needs to be generalised: the elements are pairs $(V,F)$ (again $F \in \Cls$ and $V \sse \var(F)$), such that for all partial assignments $\vp$ with $\var(\vp) \sse \var(F) \sm V$ holds $\vp * F \in \Usat \Ra \rk_1(\vp * F) = \set{\bot}$. That is, the partial assignments considered are restricted to variables not using $V$; in the same way we obtain $\ex\Urefc_k, \ex\Propc_k, \ex\Wrefc_k$. Obviously we have for all these cases $\mc{C}[\ex] \subset \ex\mc{C} \subset \ex\Up$.

CNF-Representations of boolean functions in $\Propc$ under the relative condition, i.e., representations via $\ex\Propc$, are also known as ``arc-consistent''; see \cite{GwynneKullmann2013GoodRepresentationsII} for more on this notion. There relativised hardness measurements, generalising the (p/w-)hardness as defined in Section \ref{sec:measurerepcomp}, are considered to capture the relative condition.

That $\Propc_k$, $\Urefc_k$ and $\Wrefc_k$ for the absolute condition and without new variables do not collapse, shows that a rich structure was hidden under the carpet of the relative condition aka arc consistency. A basic difference between relative and absolute condition is that under the relative condition the new variables can be used to perform certain ``computations'', since there are no conditions on the new variable other than not to distort the satisfying assignments. This is used to show the collapse to arc-consistency, as discussed in the following subsection, by encoding the stronger condition into the clause-sets in such a way that unit-clause propagation can perform the ``computations''.

\subsubsection{Separations under the relative condition}
\label{sec:seprel}

By definition it is clear that $\Propc_0$ under the absolute and under the relative conditions still just represents only the constant $0/1$ functions, and so $\Propc_0[\ex]$ as well as $\ex\Propc_0$ have modulo simple transformations just the same power as $\Propc_0$. And by Lemma \ref{lem:characrep} for $\Urefc_0 = \Wrefc_0$ under the absolute and under the relative condition we just get the same power as $\Urefc_0 = \Wrefc_0$ via equivalence, and so $\ex\Urefc_0$ and $\Urefc_0[\ex]$ have modulo simple transformations the same power as $\Urefc_0$. In \cite{GwynneKullmann2013GoodRepresentationsIII} we show that for the relative condition we have a collapse of $\Wrefc_k$ for $k \ge 1$ to $\Propc_1$ by polytime transformation (for fixed $k$), that is, all classes $\ex\Propc_k, \ex\Urefc_k, \ex\Wrefc_k$ for $k \ge 1$ can be translated in polynomial time to $\ex\Propc_1$. Thus we can summarise: under the relative condition all $\Propc_k, \Urefc_k, \Wrefc_k$ collapse to one of $\Propc_0 \subset \Urefc_0 \subset \Propc_1$ (w.r.t.\ polysize representation of boolean functions).

Motivated by \cite{BKNW2009CircuitComplexity}, in \cite{GwynneKullmann2013GoodRepresentationsII} (Theorem 6.1) we show a close connection between representations in $\ex\Urefc_1$ and monotone circuits. This leads in \cite{BeyersdorffGwynneKullmann2013PHPER} to the following separation between $\Cls$ and $\ex\Urefc_1$ (and thus, by \cite{GwynneKullmann2013GoodRepresentationsIII}, between $\Cls$ and $\ex\Wrefc_k$ for any $k$):
\begin{lem}\label{lem:sepPHP}
  For $m \in \NN$ consider the satisfiable Pigeonhole clause-sets $\php^m_m$ (``$m$ pigeons into $m$ holes''). Every sequence in $\ex\Urefc_1$ equivalent to $(\php^m_m)_{m \in \NN}$ is of superpolynomial size (in $m$ or $\ell(\php^m_m)$).
\end{lem}
Thus every representation under the relative condition of $(\php^m_m)_{m \in \NN}$ in $\Wrefc_k$ is super-polynomial, for every fixed $k$. Another example for this separation was shown in \cite{GwynneKullmann2013GoodRepresentationsII}, namely that systems of linear equations over the two-element field, in other words, systems of XOR-constraints, have obvious and short CNF-representations (in $\ex\Up$, as usual), but have no polysize arc-consistent representations. In the other direction we will see a separation in Lemma \ref{lem:xorabsc}.

The above collapse of the hierarchies under the relative condition is due to the free use of the auxiliary variables; on the contrary, under the absolute condition apparently we are very restricted with the free variables, and thus we conjecture, that there is a sequence of boolean functions which has polysize arc-consistent representations, but no polysize representations of bounded hardness, even for the w-hardness:
\begin{conj}\label{con:relhdstrong}
  There exists a polysize $(F_n)_{n \in \NN}$ in $\Cls$, with a polysize representation in $\ex\Propc$, while for no $k \in \NNZ$ there is a polysize CNF-representation $(F_n'')_{n \in \NN}$ of $(F_n)_{n \in \NN}$ in $\Wrefc_k$ under the absolute condition (i.e., in $\Wrefc_k[\ex]$).
\end{conj}

Despite of the collapse to the first level, it might be interesting to consider the classes  $\ex\Propc_k, \ex\Urefc_k, \ex\Wrefc_k$ for higher $k$, since the transformations to $\ex\Propc$ are rather costly. And perhaps a more detailed picture is revealed when considering relations more fine-grained than just using ``polysize''. We now consider the absolute condition, where we expect that all hierarchies are strict.

\subsubsection{Separations under the absolute condition}
\label{sec:sepabs}

First, to demonstrate the power of new variables is easy. We have already seen that level zero, i.e., $\Propc_0, \Urefc_0, \Wrefc_0$, does not profit from new variables. But already with the smallest class of level $1$ we can represent boolean functions which have no short CNF or DNF representations at all without new variables (as the default from now on, using the most stringent condition, the absolute condition; in KC-terminology we speak about $\Propc[\ex]$):
\begin{lem}\label{lem:xorabsc}
  The boolean function $f_n$ given by $v_1 \oplus \dots \oplus v_n = 0$, $n \in \NN$, has precisely one equivalent DNF and one equivalent CNF, each containing $2^{n-1}$ clauses of length $n$. While via splitting into sums containing only $2$ variables each, we obtain the well-known $F := X_1(v_1 \oplus \dots \oplus v_n = 0)$, as defined in \cite{GwynneKullmann2013GoodRepresentationsII}, where it is shown that $F$ is a CNF-representation of $f_n$ with $F \in \Propc$ (in other words, we got a short representation of $f_n$ in $\Propc_1[\ex]$).
\end{lem}

We now strengthen the relation $\mc{C}' \nosim \mc{C}$ between classes of clause-sets to use auxiliary variables under the strong condition (only) on the left side (note that this yields a stronger non-simulation condition than when allowing auxiliary variables on both sides):
\begin{defi}\label{def:nosima}
  For $\mc{C}, \mc{C}' \sse \Cls$ the relation \bmm{\mc{C}' \nosima \mc{C}} holds if there is a sequence $(F_n)_{n \in \NN}$ in $\mc{C}$ such that $n(F_n) = n$ and $F_n$ is computable in time $n^{O(1)}$, and such that there is no CNF-representation $(F_n')_{n \in \NN}$ of $(F_n)_{n \in \NN}$ in $\mc{C}'$ with $\ell(F_n') = n^{O(1)}$.
\end{defi}
By definition $\mc{C}' \nosima \mc{C}$ implies $\mc{C}' \nosim \mc{C}$. We are now able to state our main conjecture in its strong form (implying Conjecture \ref{con:noauxstr}):
\begin{conj}[Main Conjecture, strong form]\label{con:noauxstrstrf}
  The relation $\mc{C} \nosima \mc{C}'$ holds for classes $\mc{C}, \mc{C}' \in \set{\Propc_k, \Urefc_k, \Wrefc_k : k \in \NNZ}$ if and only if $\mc{C}' \not\sse \mc{C}$.
\end{conj}

\subsection{Knowledge compilation}
\label{sec:concKC}

We have three levels of clausal KC-formalisms:
\begin{enumerate}
\item $\Propc_k$, $\Urefc_k$ and $\Wrefc_k$ for $k \ge 0$, representing boolean functions by equivalent clause-sets in these classes; the internal relationships between these classes concerning KC (and polysize-representations) are completely covered by Conjecture \ref{con:noauxstr} (see Subsection \ref{sec:concfsep}).
\item $\Propc_k[\ex]$, $\Urefc_k[\ex]$ and $\Wrefc_k[\ex]$ for $k \ge 0$, representing boolean functions by clause-sets in these classes with existentially quantified auxiliary variables, i.e., employing the absolute condition; the internal relationships between these classes concerning KC are completely covered by Conjecture \ref{con:noauxstrstrf}.
\item $\ex\Propc_k$, $\ex\Urefc_k$ and $\ex\Wrefc_k$ for $k \ge 0$, representing boolean functions by clause-sets with existentially quantified auxiliary variables, where the defining conditions for these classes are used for the variables of the boolean function (only), i.e., employing the relative condition; the internal relationships between these classes concerning KC have been completely determined in Subsection \ref{sec:seprel}.
\end{enumerate}
The existential closure $\mc{C}[\ex]$, i.e., representing boolean functions $f$ by $F \in \mc{C}$ with $\var(f) \sse \var(F)$, and thus employing the absolute condition, has been introduced in \cite{FargierMarquis2008Closure} for KC, and further studied in \cite{Marquis2011Closures}. It has the advantage that it is easily defined for all $\mc{C}$. In contrast, the construction $\ex\Urefc$, defined in \cite{BordeauxJanotaMarquesSilvaMarquis2012UC}, apparently can not be defined as $\ex\mc{C}$ for arbitrary $\mc{C}$: a boolean function $f$ is represented by a clause-set $F$ such that for the variables of $f$ the ``underlying property'' of $\mc{C}$ holds (thus employing the relative condition).\footnote{In \cite{BordeauxJanotaMarquesSilvaMarquis2012UC} the class $\Urefc$ is called URC-C, and $\Propc$ is called UPC-C.} Extending Figure \ref{fig:introkcclassqueries}, in Figure \ref{fig:introkcclassqueries_add} the queries supported by the stronger classes $\Urefc[\ex]$ and $\ex\Urefc$ are shown. Recall that $\ex\Urefc$ can be transformed to $\ex\Propc$, which was already shown in \cite{BBCGKV2013Propc}. We see that the possibly smaller class $\Urefc[\ex]$ does not offer advantages here (recall Conjecture \ref{con:relhdstrong}). More research is needed to determine whether the absolute condition might offer some other definitive advantages for KC, for example w.r.t.\ compilation. However for SAT solving the absolute condition is superior to the relative condition, as argued in \cite{GwynneKullmann2013GoodRepresentationsII,GwynneKullmann2013GoodRepresentationsIII}, since with the absolute condition also assigning to the auxiliary variables does not lead to hard unsatisfiable problems.

Note that for the classes $\EUrefc, \UrefcE$, queries such as SE are no longer poly-time decidable. The reason for this in a nutshell is, that for CNF clause-set $F, F'$ we have $F \models F'$ iff $\fa C \in F': F \models C$, which is not the case for existentially quantified CNFs. The point is that we want implication \emph{only} on the original (free) variables, not on all variables (which we could check). In particular, it is shown in \cite{BordeauxJanotaMarquesSilvaMarquis2012UC} that $\EUrefc$, as well as other query classes built by taking the closure of $\Urefc$ under disjunction, do not allow poly-time VA, IM, EQ or SE queries (as shown in Figure \ref{fig:introkcclassqueries_add}), while $\Urefc$, and now more generally $\Urefc_k$, does.

\begin{figure}[H]
\newcommand{\thickline}{\Xhline{3\arrayrulewidth}}
  \centering
  \begin{tabular}{|c|c|c|c|c|c|c|c|c|}
    \hline
    \textbf{L} & \textbf{CO} & \textbf{VA} & \textbf{CE} & \textbf{IM} & \textbf{EQ} & \textbf{SE} & \textbf{CT} & \textbf{ME} \\ \hline\hline
    $\EUrefc$  & \tick & $\circ$ & \tick & $\circ$ & $\circ$ & $\circ$ & $\circ$ & \tick \\ \hline
    $\UrefcE$  & \tick & $\circ$ & \tick & $\circ$ & $\circ$ & $\circ$ & $\circ$ & \tick\\ \hline
  \end{tabular}
  \caption{Completing Figure \ref{fig:introkcclassqueries}, by the results for $\EUrefc$ and $\UrefcE$ from \cite{BordeauxJanotaMarquesSilvaMarquis2012UC}.}
  \label{fig:introkcclassqueries_add}
\end{figure}

In this report we concentrated on classes inside CNF ($\Cls$). In \cite{BordeauxJanotaMarquesSilvaMarquis2012UC} it is claimed (Proposition 4), that $\ex\Urefc$ simulates DNNF, citing \cite{BarahomaJungKatsirelosWalsh2008EncodingDNNF}, but there is a mistake in \cite{BarahomaJungKatsirelosWalsh2008EncodingDNNF} in that it claims that the Tseitin translation of \emph{all} DNNF's maintains arc-consistency via UCP (that is, yields $\ex\Propc$), where in fact this is only shown for smooth DNNF's as confirmed by George Katsirelos via e-mail in January 2012; so the relation of $\ex\Urefc$ (or $\ex\Propc$, which is the same here) to DNNF seems still open. Regarding the absolute condition, in \cite{GwynneKullmann2013GoodRepresentationsIII} we show that DNF can be translated in linear time to $\Propc[\ex]$.

\bibliographystyle{plain}

\newcommand{\noopsort}[1]{}

\end{document}